\documentclass[a4paper,11pt]{article}
 \usepackage{jheppub}
\usepackage{amsmath,bm,amssymb,amsthm,mathrsfs,mathtools,multirow}
\usepackage{subfigure,color}
\usepackage{ragged2e}
\justifying\let\raggedright\justifying

\allowdisplaybreaks[4]
\usepackage{slashed}
\usepackage{orcidlink}
\usepackage[table]{xcolor}
\usepackage{colortbl}
\usepackage{float}

\definecolor{headgray}{RGB}{220,220,220} 
\definecolor{blockgray}{RGB}{235,235,235} 
\definecolor{blockblue}{RGB}{220,228,242} 
\definecolor{blockorange}{RGB}{242,226,213} 
\definecolor{blockpink}{RGB}{255,230,230} 
\definecolor{blockgreen}{RGB}{220,235,210} 
\definecolor{blockgreenhi}{RGB}{190,220,175} 
\definecolor{blockyellow}{RGB}{240,232,200} 
\definecolor{blockyellowhi}{RGB}{245,225,140}
\definecolor{blockpurple}{RGB}{235,228,242}
\definecolor{blockpurplehi}{RGB}{214,198,228}

\usepackage{tikz}
\usetikzlibrary{backgrounds}
\usetikzlibrary{shapes,matrix,trees}
\usetikzlibrary{arrows.meta}
\usetikzlibrary{positioning}			
\usetikzlibrary{calc,through}			
\usetikzlibrary{decorations.pathreplacing}    
\usepackage{pgffor}                                     
\usetikzlibrary{decorations.pathmorphing}	
\usetikzlibrary{decorations.markings}
\tikzset{
	vector/.style={decorate, decoration={snake}, draw},
	provector/.style={decorate, decoration={snake,amplitude=2.5pt}, draw},
	antivector/.style={decorate, decoration={snake,amplitude=-2.5pt}, draw},
	fermion/.style={draw=black, postaction={decorate},
		decoration={markings,mark=at position .55 with {\arrow[draw=black]{>}}}},
	fermionbar/.style={draw=black, postaction={decorate},
		decoration={markings,mark=at position .55 with {\arrow[draw=black]{<}}}},
	fermionnoarrow/.style={draw=black},
	gluon/.style={decorate, draw=black,
		decoration={coil,amplitude=4pt, segment length=5pt}},
	scalar/.style={dashed,draw=black, postaction={decorate},
		decoration={markings,mark=at position .55 with {\arrow[draw=black]{>}}}},
	scalarbar/.style={dashed,draw=black, postaction={decorate},
		decoration={markings,mark=at position .55 with {\arrow[draw=black]{<}}}},
	scalarnoarrow/.style={dashed,draw=black},
	electron/.style={draw=black, postaction={decorate},
		decoration={markings,mark=at position .55 with {\arrow[draw=black]{>}}}},
	bigvector/.style={decorate, decoration={snake,amplitude=4pt}, draw},
	photon/.style={decorate, draw=black,decoration={snake,amplitude=4pt, segment length=5pt} }
}

\definecolor{ccblue}{rgb}{0.0,0.4,0.8}
\usepackage[]{hyperref}
\hypersetup{  colorlinks=true,
	linkcolor=ccblue,
	urlcolor=ccblue,
	citecolor=ccblue}

\usepackage{xcolor}

\newcommand\blfootnote[1]{
  \begingroup
  \renewcommand\thefootnote{}\footnote{#1}
  \addtocounter{footnote}{-1}
  \endgroup
}

\title{Leptogenesis Determined By Low Energy Parameters}

\author[a,b,1]{Xiao-Gang He\orcidlink{0000-0001-7059-6311}}
\author[a,b,2]{Zhong-Lv Huang\orcidlink{0009-0002-6563-4736}}
\author[c,3]{Raymond R. Volkas\orcidlink{0000-0002-4254-8520}}
\author[a,b,4]{Yu-Qi Xiao\orcidlink{0000-0002-9566-7312}}

\affiliation[a]{State Key Laboratory of Dark Matter Physics, \\
Tsung-Dao Lee Institute and School of Physics and Astronomy, \\
Shanghai Jiao Tong University,\\
1 Lisuo Road, Shanghai 201210, China}
\affiliation[b]{Shanghai Key Laboratory for Particle Physics and Cosmology, \\
Key Laboratory for Particle Astrophysics and Cosmology (MOE),\\
School of Physics and Astronomy, Shanghai Jiao Tong University,\\
800 Dongchuan Road, Shanghai 201210, China}
\affiliation[c]{ARC Centre of Excellence for Dark Matter Particle Physics, \\
School of Physics, The University of Melbourne, \\
Victoria 3010, Australia 
\blfootnote{1 hexg@sjtu.edu.cn}
\blfootnote{2 huangzhonglv@sjtu.edu.cn, corresponding author}
\blfootnote{3 raymondv@unimelb.edu.au}
\blfootnote{4 sjtu7352716@sjtu.edu.cn, corresponding author}}

\abstract{
We study thermal leptogenesis in three predictive type-I seesaw models in which the neutrino Dirac mass matrix is equal to the mass matrix of up-type quarks, or down-type quarks, or charged leptons. In this framework, the seesaw relation permits a full reconstruction of the heavy right-handed neutrino mass matrix from low-energy neutrino parameters, which greatly reduces the parameter freedom. A systematic numerical scan based on density matrix Boltzmann equations is performed to examine whether the observed baryon asymmetry of the Universe can be obtained.
Successful leptogenesis occurs for normal ordering of light neutrino masses with nonzero Majorana phases. In this case, viable solutions are found in model B, associated with down-type quarks, and model C, associated with charged leptons. Both point to a close-mass pair of heavy neutrinos satisfying $|M_i-M_j|/M_i<10^{-3}$, while remaining outside the conventional quasi-degenerate resonant regime. Four representative benchmark points are selected to show the evolution of the asymmetry and the impact of different treatments of spectator effects.
Neutrinoless double beta decay is further studied for all parameter points that can generate an acceptable baryon asymmetry $\eta_B = (6.12 \pm 0.20)\times 10^{-10}$. The predicted effective Majorana mass for certain cases can be probed by next generation experiments with sub-10 meV sensitivity, such as LEGEND-1000, nEXO, JUNO 50 tons, and CUPID-1T. This framework therefore provides clear targets for future searches.
}

\arxivnumber{}

\begin{document}
\maketitle

\section{Introduction}
\label{sec:intro}

The origin of the baryon asymmetry of the Universe (BAU) is one of the major open questions in particle physics and cosmology. Cosmological observations indicate a tiny but nonzero baryon-to-photon ratio~\cite{Planck:2018vyg}, given by
\begin{align}
\eta_B \equiv \frac{n_B-n_{\bar B}}{n_\gamma} \simeq (6.12\pm0.04)\times 10^{-10}~.
\end{align}
Generating this asymmetry from an initially baryon-symmetric Universe requires the three Sakharov conditions~\cite{Sakharov:1967dj}: baryon number violation, C and CP violation, and departure from thermal equilibrium.
Although the Standard Model (SM) contains anomalous baryon number violation and CP violation, the resulting asymmetry is not sufficient for the observed value~\cite{Kuzmin:1985mm,Harvey:1990qw,Gavela:1993ts}. New physics beyond the SM is therefore required, and many extensions of the SM have been proposed to address this problem~\cite{Yoshimura:1978ex,Affleck:1984fy,Fukugita:1986hr,Turok:1990zg,Carena:1996wj,Dine:2003ax,Davidson:2008bu}.

Leptogenesis is particularly attractive because it can account for the BAU while relating it to the origin of tiny neutrino
masses~\cite{Fukugita:1986hr,Barbieri:1999ma,Buchmuller:2004nz,Davidson:2008bu}. In the simplest realization, the Type-I seesaw model~\cite{Minkowski:1977sc,Gell-Mann:1979vob,Yanagida:1979as,Mohapatra:1979ia}, tiny Majorana neutrino masses arise from heavy right-handed (RH) Majorana neutrinos. Their CP violating decays generate a lepton asymmetry, or equivalently a nonzero $B-L$ asymmetry, which can be partially converted into a baryon asymmetry by electroweak sphaleron processes~\cite{Kuzmin:1985mm,Harvey:1990qw}.
In the Type-I seesaw model, the Lagrangian can be written as
\begin{align}
-\mathcal{L}\supset
\overline{L_{L,\alpha}}(Y_e)_{\alpha}\Phi E_{R,\alpha}
+h_{\alpha i}\overline{L_{L,\alpha}}\widetilde{\Phi}N_{R,i}
+\frac{1}{2}\overline{N_{R,i}^{c}}(M_R)_{ij}N_{R,j}
+\mathrm{h.c.}~,
\label{eq:Lag-seesaw}
\end{align}
 in the basis where the charged lepton Yukawa matrix $Y_e$ is diagonal but the heavy RH Majorana neutrino mass matrix $M_R$ is generally not. Here, $\alpha$ and $i,j$ are flavor indices, $L_{L,\alpha}=(\nu_{L,\alpha},e_{L,\alpha})^T$ are the lepton doublets, $E_{R,\alpha}\equiv e_{R,\alpha}$ represent the charged lepton singlets, $N_{R,i}$ denote the heavy RH Majorana neutrinos, and the Higgs doublet is written as $\Phi=(\phi^{+},\phi^{0})^{T}$, where $\widetilde\Phi=i\sigma_2\Phi^*$ and $\sigma_2$ is the second Pauli matrix. After electroweak symmetry breaking, the Higgs doublet acquires the vacuum expectation value (VEV) $\langle \Phi\rangle=v/\sqrt{2}$ with $v=246~\mathrm{GeV}$. The neutrino Yukawa coupling matrix $h$ is then related to the neutrino Dirac mass matrix by $\bar{m}^D_{\nu}=h v/\sqrt{2}$ in the diagonalized $Y_{e}$ basis. In the following discussions, $m_{\nu}^D$ denotes the neutrino Dirac mass matrix in a \textit{general} flavor basis. The matrix $M_R$ is diagonalized before the leptogenesis calculation is performed. 

There is usually a large parameter space in the Type-I seesaw model that can accommodate the BAU through leptogenesis over a wide range of heavy neutrino masses. The matrices $M_{R}$, and $\overset{\scriptscriptstyle(-)}{m}~\!\!_{\nu}^D$ introduce new CP violating sources, which may be sufficient to explain the BAU when heavy neutrinos decay and sphaleron processes are effective.
However, they contain many free parameters that cannot be directly measured at low energies~\cite{Casas:2001sr,Davidson:2008bu}. As a result, the framework is not fully predictive, making it difficult to directly connect neutrino oscillation data with the high energy physics relevant for leptogenesis. It is desirable to find ways to determine these new CP violating sources in $M_R$, and $\overset{\scriptscriptstyle(-)}{m}~\!\!_{\nu}^D$ from low energy observables in neutrino physics.

In previous papers, the neutrino Dirac mass matrix $m^D_{\nu}$ was usually related to the up-type quark mass matrix through $SO$(10)-inspired conditions~\cite{Buchmuller:1996pa,Nezri:2000pb,Buccella:2001tq,Branco:2002kt,Akhmedov:2003dg,DiBari:2008mp}, or $m_{\nu}^D\propto I $ was induced by a flavor symmetry and thus a certain texture for the $M_{R}$ matrix was required~\cite{Smirnov:1993af,Ma:2001dn,Hagedorn:2006ug,Altarelli:2010gt,King:2013eh,Ding:2024ozt,Ma:2026ivz}. There is, however, another well-motivated way as shown in~\cite{He:2008cd}: predictive flavor or unified constructions in UV-complete models can naturally determine $m^D_{\nu}$ through the measured charged fermion spectrum. In addition, $M_R$ and $m_{\nu}^D$ are related through the seesaw formula,
\begin{align}
m_\nu \approx - m_\nu^D \,M^{-1}_R\,(m_\nu^{D})^T~, \label{seesaw}
\end{align}
where $m_\nu$ is the light-neutrino mass matrix. It can be diagonalized by the Pontecorvo--Maki--Nakagawa--Sakata (PMNS) matrix as~\cite{Maki:1962mu}
\begin{align}
    m_\nu = U_{\rm PMNS}\, \hat m_\nu\, U^T_{\rm PMNS}~, \label{PMNS}
\end{align}
where the charged leptons are in the mass basis and $\hat m_\nu ={\rm{Diag}}(m_1, m_2, m_3)$, with $m_i$ denoting the light-neutrino masses. These relations reduce the number of degrees of freedom in the parameter space.

In this work, we study leptogenesis in a class of predictive seesaw models in which $m_{\nu}^D= K \hat{m}_{f}$, where $\hat{m}_f$ is the diagonal matrix of mass eigenvalues for charged-fermion $f$ while $K$ is a known matrix of numbers, and thus $M_R$ can be determined by low energy neutrino parameters. The main purpose is to examine whether these models can generate the observed BAU through leptogenesis. Three specific models were proposed in Ref.~\cite{He:2008cd}, and they lead to
\begin{align}
M_R \simeq \hat m_f\, U_{\rm PMNS}^*\, \hat m_\nu^{-1}\, U_{\rm PMNS}^\dagger\, \hat m_f~,
\qquad f=e~,d~,u~.
\label{eq:MR-general}
\end{align}
The three specific models are labeled as follows: 
\begin{align*}
{\rm{\bf Model~A:}}~f=u~; \quad
{\rm{\bf Model~B:}}~f=d~; \quad
{\rm~and~{\bf Model~C:}}~f=e~.
\end{align*}
Once the relation in Eq.~\eqref{eq:MR-general} is imposed, $M_R$ and $m_{\nu}^D$ are no longer arbitrary. In this case, the neutrino sector inputs are reduced to three neutrino oscillation angles $\theta_{ij}$ and the mass squared differences $\Delta m_{ij}$, which can be experimentally determined, and four other free parameters: the lightest neutrino mass $m_{\rm lightest}$, the Dirac phase $\delta_{\text{CP}}$, and two Majorana phases $\alpha$ and $\beta$. This makes the framework much more predictive. Rather than concentrating on the model building details, which are summarized in Sec.~\ref{sec:modelsreview}, we focus on their phenomenological implications for leptogenesis as our main purpose.

To compute the final $B-L$ asymmetry $N_{B-L}^{\rm f}$, we assume a vanishing pre-existing asymmetry and solve the Boltzmann equations in the density matrix formalism following Ref.~\cite{Blanchet:2011xq}, as shown in Sec.~\ref{sec:seesaw}. These equations contain the CP violating source, decay, and washout terms and include spectator effects as discussed in Refs.~\cite{Davidson:2008bu,Antusch:2010ms}. The final baryon-to-photon ratio $\eta_B$ can be related to $N_{B-L}^{\rm f}$ by the sphaleron conversion factor. This method naturally accounts for quantum decoherence, flavor effects, and gauge interactions in the analysis.

A detailed numerical analysis is shown in Sec.~\ref{sec:numerics}. It includes the numerical inputs and their uncertainties, representative benchmark points, the effect of finite rate spectator equilibration, and parameter scans consistent with neutrino oscillation data. These results are used to test whether the three specific models can reproduce the observed baryon asymmetry, and to derive their predictions for the effective neutrino mass relevant to neutrinoless double beta decay. Finally, Sec.~\ref{sec:conclusion} provides our conclusions.

\section{Brief review of the underlying models}
\label{sec:modelsreview}

The goal is to obtain Eq.~(\ref{eq:MR-general}) for the three different cases $f=e,d,u$ while avoiding phenomenologically unacceptable constraints on quark and lepton masses and mixing matrices. The technical model-building prescriptions are fully described in Ref.~\cite{He:2008cd}. We refrain from recapitulating all of those details here, being content to summarize the main points for completeness.\footnote{Note that Ref.~\cite{He:2008cd} provides candidate models so as to constitute existence proofs. Other models are also expected to exist.}

There are two general ingredients: a symmetry-induced relation between the neutrino Dirac mass matrix and the chosen charged-fermion masses, together with restrictions on various fermion diagonalization matrices. The former may be achieved by utilizing some form of quark-lepton symmetry for the $f=d,u$ cases, while right-handed weak isospin can be used for the $f=e$ scenario. The diagonalization matrix restrictions require a flavor symmetry.

\vspace{3mm}

\noindent
\textbf{Model A:}\ \ The flipped $SU(5) \times U(1)$ gauge symmetry may be used to relate $m_\nu^D$ to $m_u$. Recall that in these models the roles of down and up antiquarks are flipped, as are the roles of charged antileptons and antineutrinos. Thus the usual $SU(5)$-type mass relation $m_d = m_e$ instead becomes $m_u = m_\nu^D$. To constrain the relevant diagonalization matrices, an $A_4$ flavor symmetry may be used. The upshot is that $m_\nu^D = - U_\omega \hat{m}_u$ is achieved at tree-level, with
\begin{equation}
    U_\omega = \frac{1}{\sqrt{3}} \left( \begin{array}{ccc}
    1 & 1 & 1 \\
    1 & \omega & \omega^2 \\
    1 & \omega^2 & \omega
    \end{array} \right)~,
\end{equation}
where $\omega= \exp(2 \pi i/3)$ is a cube root of unity. 
The model also produces $m_e = U_\omega \hat m_e$, meaning that the left-diagonalization matrix, $V_{eL}$, for charged leptons obeys $V^\dagger_{eL}=U_\omega$. Equation~(\ref{seesaw}) defines the neutrino diagonalization matrix $V_\nu$ through $\hat{m}_\nu = V_\nu m_\nu^D M_R^{-1} (m_\nu^D)^T V^T_\nu$, so that $U_{\rm{PMNS}} \equiv V_{eL} V^\dagger_\nu = U^\dagger_\omega V^\dagger_\nu$. This immediately leads to Eq.~(\ref{eq:MR-general}) with $f = u$. The PMNS and CKM mixing matrices are unrestricted.

\vspace{3mm}

\noindent
\textbf{Model B:}\ \ For the $f=d$ case one may use the discrete quark-lepton symmetry~\cite{Foot:1990dw,Foot:1990um,Foot:1991fk,Foot:1995xx} idea rather than a more familiar continuous symmetry. The structure of the model relates $d$-type quarks and neutrinos, as well as $u$-type quarks and charged leptons. The gauge group is $SU(3)_\ell \times SU(3)_q \times SU(2)_L \times U(1)_X$, where $SU(3)_q$ is the usual QCD while $SU(3)_\ell$ is \textit{leptonic color}. A $Z_2$ discrete $q \leftrightarrow \ell$ symmetry may be imposed to relate quark multiplets to generalized lepton multiplets. Leptonic color is spontaneously broken, with standard leptons identified as one of the three leptonic colors. The other two leptonic colors become heavy exotic fermions. As for Model A, an $A_4$ flavor symmetry is used to restrict the diagonalization matrices. The result is that $m_\nu^D = U_\omega \hat{m}_d$ and $m_e = U_\omega \hat{m}_e$ may be obtained. Renaming $M_R$ as the negative of the original $M_R$, one identifies $f = d$. There are no constraints on the up-type quark and charged-lepton mass matrices. The PMNS matrix is also unrestricted, while at tree-level the CKM matrix is the identity. Higher-order corrections will induce nonzero CKM mixing, though the details remain to be worked out.

\vspace{3mm}

\noindent\textbf{Model C:}\ \ Left-right symmetric models, based on the gauge group $SU(3)_c \times SU(2)_L \times SU(2)_R \times U(1)_{B-L}$, have the power to relate the fermion masses of weak-isospin partners, which is what neutrinos and charged leptons are. This is done by utilizing a \textit{real} bidoublet Higgs multiplet in place of the more common complex bidoublet (see, for example, Ref.~\cite{Volkas:1995yn}). To relate $m_\nu^D$ with $m_e$ while avoiding any relation between $m_u$ and $m_d$, one uses a real bidoublet in the lepton sector but a complex bidoublet in the quark sector. As above, the flavor symmetry is $A_4$. The outcome is that $m_\nu^D = - U_\omega \hat{m}_e$ and $m_e = U_\omega \hat m_e$ are obtained, leading to $f = e$ with no restrictions on the quark masses or the PMNS and CKM mixing matrices.

\section{Framework for Seesaw Leptogenesis}
\label{sec:seesaw}

This section introduces the notation and main equations for the leptogenesis analysis. The discussion proceeds in the basis where the charged lepton mass matrix is diagonal, while the RH neutrino mass matrix is diagonalized by a unitary transformation. The flavored CP asymmetries and the density matrix evolution equations, including spectator effects, are presented. The sphaleron conversion of the lepton asymmetry into baryon asymmetry is also briefly summarized. The framework described below follows Ref.~\cite{Blanchet:2011xq}.

\paragraph{CP Asymmetry}
As mentioned above, it is convenient to work in the basis in which the charged lepton mass matrix is diagonal, while $M_R$ is diagonalized by a unitary matrix $U_R$,
\begin{align}
U_R^T\,M_R\,U_R = \hat M_R \equiv {\rm Diag}(M_1,M_2,M_3)~,
\end{align}
where the masses are ordered as $0<M_1\leq M_2\leq M_3$. In this basis, the $\bar{m}^D_\nu$ matrix and the corresponding Yukawa coupling $h$ in Eq.~\eqref{eq:Lag-seesaw} are rotated as
\begin{align}
\tilde{m}^D_{\nu} = \bar{m}^D_\nu\,U_{R}~,\quad
\tilde{h} =  h\,U_{R}~.
\label{eq:mD-lepto-basis}
\end{align}
The CP violating source is controlled by the flavored CP asymmetry matrix for each heavy neutrino decay $N_i \to L_{L,\alpha} \Phi$ and the corresponding antiparticle decay $N_i \to \overline{L_{L,\alpha}}\Phi^{\dagger}$, which are assumed to be the dominant processes generating the asymmetry.
The flavored CP asymmetry matrix can be expressed as~\cite{Blanchet:2011xq}
\begin{align}
\varepsilon^{(i)}_{\alpha\beta}
=
\dfrac{3}{32\pi\,(\tilde{h}^\dagger \tilde{h})_{ii}}
\sum_{j\neq i}
\bigg\{
&i\,[\tilde{h}_{\alpha i}\tilde{h}^*_{\beta j}(\tilde{h}^\dagger \tilde{h})_{ji}
-\tilde{h}^*_{\beta i} \tilde{h}_{\alpha j}(\tilde{h}^\dagger \tilde{h})_{ij}]\,
\dfrac{\xi(x_{ji})}{\sqrt{x_{ji}}}\notag\\
+&\dfrac{2i}{3(x_{ji}-1)}\,
[\tilde{h}_{\alpha i} \tilde{h}^*_{\beta j}(\tilde{h}^\dagger \tilde{h})_{ij}
-\tilde{h}^*_{\beta i}\tilde{h}_{\alpha j}(\tilde{h}^\dagger \tilde{h})_{ji}]
\bigg\}~,
\label{eq:eps-matrix}
\end{align}
where $x_{ji}\equiv {M_j^2}/{M_i^2}$, and the function $\xi(x)$ is defined as
\begin{align}
\xi(x)=\frac{2}{3}x\left[(1+x)\ln\!\left(\frac{1+x}{x}\right)-\frac{2-x}{1-x}\right]~.
\label{eq:loopfunction}
\end{align}
The diagonal entries reproduce conventional flavored CP asymmetries $\varepsilon^{(i)}_{\alpha\alpha}=\epsilon_{i\alpha}$ as discussed in~\cite{Abada:2006ea,Nardi:2006fx}, while the off-diagonal entries are needed in the density matrix treatment. In this work, we do not consider the exactly degenerate limit $x_{ji}\to 1$, where a proper resonant treatment would be required. 

\paragraph{Boltzmann Equation}
In the density matrix formalism, the $B-L$ asymmetry is represented by a $3\times 3$ matrix in the charged lepton mass basis. Its diagonal entries describe the flavor asymmetries, while the off-diagonal entries encode flavor coherence. The abundance of the heavy RH Majorana neutrinos $N_{i}$ satisfies the evolution equations~\cite{Blanchet:2011xq}
\begin{align}
\frac{dN_{N_i}}{dz}
=
- \dfrac{M_i}{M_1}\, D_i(z)\,
\left(N_{N_i}-N_{N_i}^{\rm eq}\right),
\label{eq:BE-RHN}
\end{align}
with a dimensionless $N_{X}\equiv n_{X}V_{c}=n_{X}/n_{N_{i}}^{\rm eq}(T\gg M_{i})$ denoting the number of $X$ particles (with $X=N_{i}$), or the amount of $B-L~\rm{charge\ asymmetry}$, in a co-moving volume $V_{c}$. As a convention, that volume is chosen such that it contains one ultra-relativistic RH neutrino in thermal equilibrium, meaning that $V_{c}n_{N_{i}}^{\rm eq}(T\gg M_{i})=1$, where $n_{X}$ is the number density. The density matrix evolves as
\begin{align}
\dfrac{dN_{B-L}^{\alpha\beta}}{dz}
&=\sum\limits_{i=1,2,3}\bigg[
\varepsilon^{(i)}_{\alpha\beta}\,
\dfrac{M_i}{M_1}D_i(z)\,
(N_{N_i}-N_{N_i}^{\rm eq})
-\dfrac{1}{2}
\dfrac{M_i}{M_1}W_i(z)\,
\{P^{0(i)},N_{B-L}^{\rm spec}\}_{\alpha\beta}\bigg]
\notag\\
&-\sum_{\ell=e,\mu,\tau}
\dfrac{{\rm Im}(\Lambda_\ell)}{H(z)z}\,
\left[\lambda_\ell,\left[\lambda_\ell,N_{B-L}\right]\right]_{\alpha\beta}~,
\label{eq:BE-density}
\end{align}
where $\{\cdot,\cdot\}$ denotes the anti-commutator and
$\lambda_e=\mathrm{Diag}(1,0,0),~
\lambda_\mu=\mathrm{Diag}(0,1,0),~
\lambda_\tau=\mathrm{Diag}(0,0,1)$. These equations are evolved simultaneously using the common temperature variable $z=M_1/T$, from the initial $z_{\rm 0}$ to the final $z_{\rm f}$. Compared with the formulation given by Ref.~\cite{Blanchet:2011xq}, these Boltzmann equations make the Jacobian factor $dz_{i}/dz=M_{i}/M_{1}$ explicit. 
The decay term $D_i$ and inverse decay washout term $W_i$ can be expressed as
\begin{align}
D_i(z)=K_i \,z \,\dfrac{M_i}{M_1}\,
\dfrac{\mathcal{K}_1[(M_i/M_1)z]}{\mathcal{K}_2[(M_i/M_1)\,z]},
\quad
W_i(z)=\dfrac{1}{4}\,K_i \,
z^3\bigg(\dfrac{M_i}{M_1}\bigg)^3\,
\mathcal{K}_1[(M_i/M_1)\,z]~,
\end{align}
where $\mathcal{K}_n$ are the modified Bessel functions, and $K_{i}$ is the total decay parameter for $N_{i}$ as defined below. Although $D_i$ and $W_i$ are usually written as functions of the species variable $z_i$, here they have been converted using $z_i=M_i/T=zM_{i}/M_{1}$ for use in the common-$z$ equations. The $K_{i}$ is defined as the ratio of the decay rate to the Hubble expansion rate
\begin{align}  
K_i \equiv\dfrac{\Gamma_{i}+\bar{\Gamma}_i}{H(T=M_i)}= \dfrac{\tilde m_i}{m_*}~,
\label{eq:Kpara}
\end{align}
where $\Gamma_i^{\rm tot}=\Gamma_{i}+\bar{\Gamma}_i=(\tilde{h}^\dagger \tilde{h})_{ii}M_{i}/8\pi$ is the tree level total decay rate, summed over both CP-conjugate channels, of $N_{i}$, $\widetilde m_i \equiv \left[(\tilde{m}_\nu^D)^\dagger \tilde{m}_\nu^D\right]_{ii}/M_i$ is the effective mass~\cite{Plumacher:1996kc}, and $m_* \simeq 1.08\times 10^{-3}\ {\rm eV}$ is the equilibrium mass~\cite{Nezri:2000pb,Buchmuller:2004nz}.
At the flavor level, the tree-level projector is defined as
$P^{0(i)}_{\alpha\beta}
= (\tilde{h}_{\alpha i} \tilde{h}_{\beta i}^{*})
/(\tilde{h}^\dagger \tilde{h})_{ii}$,
so that the flavored effective masses can be written as $\widetilde m_{i\alpha}=P^{0(i)}_{\alpha\alpha}\,\widetilde m_i$ and decay parameters are
\begin{align}
K_{i\alpha}=P^{0(i)}_{\alpha\alpha}\,K_i~,\quad \sum\limits_{\alpha=e,\mu,\tau}K_{i\alpha}=K_{i}~.
\label{eq:K}
\end{align}
According to the definition of the $N_{X}$, the equilibrium abundance of $N_i$ can be written as
\begin{align}
N^{\rm eq}_{N_i}(z)
=\dfrac{1}{2}\, z^2 \,
\bigg(\dfrac{M_i}{M_1} \bigg)^2\,
\mathcal{K}_2(z_i)~.
\label{eq:abundance}
\end{align}

The last term in Eq.~\eqref{eq:BE-density} accounts for flavor decoherence, where $\Lambda_\ell$ denotes the charged lepton self-energy correction and its imaginary part is approximated by ${\rm Im}(\Lambda_\ell)\simeq 8\times10^{-3} f_\ell^2 T$~\cite{Cline:1993bd,Blanchet:2011xq}, with $f_{\ell}=Y_{e,\ell}(\Lambda)$ including the RG running effects up to the $10^{9}$ GeV scale.
The Hubble rate is expressed in terms of the common variable $z$ as $H(z)=1.66\sqrt{g_*}(M_1/z)^2/M_{\rm Pl}$ where $g_* = 106.75$ denotes the effective number of relativistic degrees of freedom in the SM at temperatures above the electroweak scale~\cite{Kolb:1990vq}. The relative size of ${\rm Im}(\Lambda_{\alpha})$ and the Hubble rate determines whether flavor decoherence is effective.

\paragraph{Spectator Effects}
Certain fast processes, known as spectator processes~\cite{Buchmuller:2001sr,Nardi:2005hs}, can affect the lepton number indirectly by changing the densities of the lepton doublets and the Higgs. In the density matrix formalism, these effects act only on the flavor diagonal part of $N_{B-L}$, since they modify the charge asymmetries in the thermal bath without directly affecting the off-diagonal part. Accordingly, $N_{B-L}$ can be replaced by
\begin{align}
N_{B-L}^{\rm spec}
=
N_{B-L}
-\mathrm{Diag}(N_{B-L})
+\mathrm{Diag}[\mathcal{C}(z)\cdot\mathrm{Diag}(N_{B-L})]~,
\end{align}
where the matrix $\mathcal{C}(z)=\mathcal{C}_{L_L}(z)+\mathcal{C}_{\Phi}$ describes how spectator effects redistribute the flavor asymmetries. Its form depends on the temperature range and includes contributions from the lepton doublets and the Higgs field. The detailed values of the matrix will be shown in Sec.~\ref{subsec:input}. As a result, spectator effects can modify both the size of the final asymmetry and its flavor composition, and are therefore important for a reliable numerical analysis.

\paragraph{Final Baryon Asymmetry}
After solving the Boltzmann equations in Eqs.~\eqref{eq:BE-RHN} and \eqref{eq:BE-density}, the final $B-L$ asymmetry is obtained from
\begin{align}
N_{B-L}^{\rm f} = {\rm Tr}\,[N_{B-L}(z_{\rm f})]~,
\end{align}
where the trace sums over the flavor components after the evolution. The corresponding the baryon-to-photon ratio $\eta_B$ is then related to $N_{B-L}^f$ through the sphaleron process~\cite{Blanchet:2011xq,Buchmuller:2004nz,Nardi:2005hs}
\begin{align}
\eta_B=a_{\rm sph}\,\dfrac{N_{B-L}^{\rm f}}{N_{\gamma}^{\rm rec}}\simeq 9.6\times10^{-3}~N_{B-L}^{\rm f}~,
\end{align}
where $N_{\gamma}^{\rm rec}\simeq 37$ denotes the number of photons at recombination~\cite{DiBari:2005st}, and $a_{\rm sph}=n_{B}/n_{B-L}$ denotes the sphaleron conversion factor from the $B-L$ asymmetry to the baryon asymmetry. The factor takes the value $a_{\rm sph}=28/79$~\cite{Harvey:1990qw,Laine:1999wv} under the assumption that the electroweak sphaleron leaves equilibrium before the electroweak phase transition.

This completes the framework for the leptogenesis analysis. The next section turns to the numerical study of the three predictive models, including representative benchmark points and the viable parameter regions.

\section{Numerical Analysis}
\label{sec:numerics}
This section presents the numerical analysis for the three models, {\bf A}, {\bf B} and {\bf C}, as described earlier. The numerical inputs and their uncertainties are described first, and the resulting parameter scans, benchmark points, and phenomenological implications are then discussed.

\subsection{Numerical Inputs and Uncertainties}
\label{subsec:input}
This subsection describes the numerical inputs used in the analysis and the main sources of uncertainty. The relevant neutrino parameters and charged fermion masses are specified first, together with the conventions adopted in the numerical calculation. The treatment of spectator effects is then summarized, followed by a brief discussion of the uncertainties that are most relevant for the prediction of the baryon asymmetry.

\paragraph{Input Parameters and Conventions}
The PMNS matrix, which depends on six independent parameters, has the standard parameterization
\begin{align}
U_{\rm PMNS}=
\begin{pmatrix}
c_{12}c_{13} & s_{12}c_{13} & s_{13}e^{-i\delta_{\rm CP}}\\
-s_{12}c_{23}-c_{12}s_{23}s_{13}e^{i\delta_{\rm CP}} &
c_{12}c_{23}-s_{12}s_{23}s_{13}e^{i\delta_{\rm CP}} &
s_{23}c_{13}\\
-s_{12}s_{23}+c_{12}c_{23}s_{13}e^{i\delta_{\rm CP}} &
c_{12}s_{23}+s_{12}c_{23}s_{13}e^{i\delta_{\rm CP}} &
-c_{23}c_{13}
\end{pmatrix}
\begin{pmatrix}
1&0&0\\
0&e^{i\alpha}&0\\
0&0&e^{i\beta}
\end{pmatrix}~,
\label{eq:UPMNS-param}
\end{align}
where $s_{ij}=\sin{\theta_{ij}}$ and $c_{ij}=\cos{\theta_{ij}}$ with $\theta_{ij}~(i,j=1,2,3)$ denoting the mixing angles which have been precisely determined by experiments. The parameter $\delta_{\rm CP}\in [0,2\pi]$ is the Dirac CP phase, and $\alpha,\beta\in [0,\pi]$ denote the two Majorana phases in our convention. For the neutrino mass ordering, both normal ordering (NO) and inverted ordering (IO) are considered in the analysis, parameterized in terms of the lightest neutrino mass $m_{\rm lightest}$ and the experimentally measured mass squared differences
\begin{align}
{\rm NO}:&\quad
m_1=m_{\rm lightest}~,&&
m_2=\sqrt{m_1^2+\Delta m_{21}^2}~,&&
m_3=\sqrt{m_1^2+\Delta m_{31}^2}~,\\
{\rm IO}:&\quad
m_3=m_{\rm lightest}~,&&
m_1=\sqrt{m_3^2+|\Delta m_{31}^2|}~,&&
m_2=\sqrt{m_3^2+|\Delta m_{32}^2|}~.
\end{align}
The NuFIT-6.1~\cite{Esteban:2024eli,nufit} best-fit values are taken for the parameters with NO (IO),
\begin{align}
\Delta m_{31(2)}^2 &= 2.521~(-2.500)\times 10^{-3}\ {\rm eV}^2~,\quad 
\Delta m_{21}^2 = 7.537\times 10^{-5}\ {\rm eV}^2~,\notag\\
\theta_{12} &= 33.76^\circ~,\qquad
\theta_{13}=8.62^\circ~(8.65^\circ)~,\qquad
\theta_{23}=43.27^\circ~(48.15^\circ)~,
\end{align}
For the charged fermion masses $\hat{m}_{f}$ entering the construction of $m_\nu^D$ and $M_R$, the $\overline{\rm MS}$ running masses around the typical leptogenesis energy scale, $\Lambda=10^{9}$ GeV, are used in the scanning analysis. The input masses are taken from Ref.~\cite{He:2008cd}:
\begin{align}
m_{u}(\Lambda)&=1.3~{\rm{MeV}}~,&&
m_{c}(\Lambda)=370~{\rm{MeV}}~,&&
m_{t}(\Lambda)=110~{\rm{GeV}}~,\\
m_{d}(\Lambda)&=2.6~{\rm{MeV}}~,&&
m_{s}(\Lambda)=52~{\rm{MeV}}~,&&
m_{b}(\Lambda)=1.5~{\rm{GeV}}~,\\
m_{e}(\Lambda)&=0.52~{\rm{MeV}}~,&&
m_{\mu}(\Lambda)=110~{\rm{MeV}}~,&&
m_{\tau}(\Lambda)=1.8~{\rm{GeV}}~.
\end{align}
Since the heavy RH neutrino masses may span several orders of magnitude, the use of running Yukawa couplings $f_{\alpha}=Y_{e,\alpha}(\Lambda)$ at the scale $\Lambda$ is also important for a quantitatively reliable determination of both the heavy spectrum and the leptogenesis parameters. 

With the neutrino mixing angles and mass squared differences fixed at their experimental best-fit values, and with the charged fermion masses specified at the leptogenesis scale as described above, the reconstructed Dirac and Majorana neutrino mass matrices are effectively determined up to the four remaining low-energy parameters $m_{\rm lightest},\delta_{\rm CP},\alpha,\beta$. Accordingly, in the present framework the final baryon asymmetry can be regarded as a function
\begin{align}
    \eta_B=\eta_B(m_{\rm lightest},~\delta_{\rm CP},~\alpha,~\beta)~.
\end{align}

For the leptogenesis analysis, the numerical evolution is performed in terms of the common variable $z=M_1/T$. Following the standard treatment of thermal leptogenesis~\cite{Davidson:2008bu,Buchmuller:2004nz}, the integration range is chosen such that $z_{\rm 0}\ll1$ and $z_f\gg1$, so that the evolution starts in the relativistic regime and ends after the final asymmetry has frozen in. In the general scan, we take $z_{\rm 0}=10^{-11}$ and $z_f=30$. Since the equilibrium abundance of each RH neutrino depends on $z_i=(M_i/M_1)z$, the small value of $z_{\rm 0}$ guarantees $z_i\ll1$ for all species, including the heaviest state $N_3$, and hence $N_{N_i}^{\rm eq}(z_{\rm 0})\simeq 1$, as can be derived from Eq.~\eqref{eq:abundance}. Assuming thermal initial conditions, $N_{N_i}(z_{\rm 0})= N_{N_i}^{\rm eq}(z_{\rm 0})$ has been set. Note again that no pre-existing asymmetry is assumed $N_{B-L}(z_{\rm 0})=0$.

In the numerical analysis, spectator effects are implemented by using a standard description~\cite{Buchmuller:2001sr,Nardi:2006fx,Antusch:2006cw}, which behaves as an instantaneous switch at $T\sim10^9~\mathrm{GeV}$, to get the main results. In the temperature range $10^{9}$--$10^{12}$ GeV, where the $\tau$ Yukawa interaction is in equilibrium but the $e$ and $\mu$ Yukawa interactions are not, the effects are described by the two-flavor spectator matrix $\mathcal{C}^{(2)}$. Usually, it can be described by a $2\times2$ matrix
\begin{align}
\mathcal{C}^{(2)}_{2\times2}= 
\mathcal{C}_{L_{L},2\times2}^{(2)}
+\mathcal{C}_{\Phi,2\times2}^{(2)}=
\begin{pmatrix}
581/589\, & 104/589 \\
134/589\, & 614/589
\end{pmatrix}~.
\end{align}
This result differs slightly from the corresponding expression in Ref.~\cite{Antusch:2010ms}, where there appears to be a minor typo. The contributions from the lepton and Higgs agree with those given in Ref.~\cite{Antusch:2010ms}:
\begin{align}
\mathcal{C}_{L_{L},2\times2}^{(2)}=
\begin{pmatrix}
417/589\, & -120/589\\
-30/589\, & 390/589
\end{pmatrix}~,\quad
\mathcal{C}_{\Phi,2\times2}^{(2)}=
\begin{pmatrix}
164/589\, & 224/589\\
164/589\, & 224/589
\end{pmatrix}~.
\end{align}
In the $3\times3$ density matrix formalism, it can be embedded in the charged lepton mass basis within the $e\leftrightarrow\mu$ symmetric matrix
\begin{align}
\mathcal{C}^{(2)}=
\mathcal{C}^{(2)}_{L_{L}}+\mathcal{C}^{(2)}_{\Phi}=
\begin{pmatrix}
667/589\, & 78/589\, & 164/589\\
78/589\, & 667/589\, & 164/589\\
134/589\, & 134/589\, & 614/589
\end{pmatrix}~,
\end{align}
where
\begin{align}
C_{L_{L}}^{(2)}=
\begin{pmatrix}
503/589\, & -86/589\, & -60/589\\
-86/589\, & 503/589\, & -60/589\\
-30/589\, & -30/589\, & 390/589
\end{pmatrix}~,~
\mathcal{C}_{\Phi}^{(2)}=
\begin{pmatrix}
164/589\, & 164/589\, & 224/589\\
164/589\, & 164/589\, & 224/589\\
164/589\, & 164/589\, & 224/589
\end{pmatrix}~.
\end{align}
At lower temperatures, where all three charged lepton flavors can be distinguished, the spectator matrix is replaced by the corresponding three-flavor form
\begin{align}
\mathcal{C}^{(3)}=
\mathcal{C}^{(3)}_{L_{L}}+\mathcal{C}^{(3)}_{\Phi}=
\begin{pmatrix}
188/179\, & 32/179\, & 32/179\\
49/358\, & 500/537\, & 142/537\\
49/358\, & 142/537\, & 500/537
\end{pmatrix}~,
\end{align}
with
\begin{align}
C_{L_{L}}^{(3)}=
\begin{pmatrix}
151/179\, & -20/179\, & -20/179\\
-25/358\, & 344/537\, & -14/537\\
-25/358\, & -14/537\, & 344/537
\end{pmatrix}~,~
\mathcal{C}_{\Phi}^{(3)}=
\begin{pmatrix}
37/179\, & 52/179\, & 52/179\\
37/179\, & 52/179\, & 52/179\\
37/179\, & 52/179\, & 52/179
\end{pmatrix}~,
\end{align}
In the common variable $z=M_1/T$, the switch therefore occurs as per
\begin{align}
\mathcal{C}(z)=
\begin{cases}
\mathcal{C}^{(2)}~,&z\leq z_{\rm switch}~,\\
\mathcal{C}^{(3)}~,&z>z_{\rm switch}~,
\end{cases}
\quad
z_{\rm switch}=\dfrac{M_1}{10^9~\mathrm{GeV}}~.
\label{eq:spectator-switch}
\end{align}
The above prescription implements spectator effects through a single instantaneous switch of the plasma response, applied uniformly to the $N_{i}$ contributions. A corresponding finite rate treatment is also discussed in Sec.~\ref{subsec:finite-spectator}.

\paragraph{Sources of Uncertainty}
In the prediction of the baryon asymmetry, the uncertainties mainly come from three sources: the experimental errors on the neutrino oscillation parameters, the charged fermion running masses, and the treatment of spectator effects in the Boltzmann evolution.

\begin{itemize}
\item {\it Neutrino oscillation inputs.} The leptonic mixing angles and mass squared differences are known experimentally, but not exactly. Their current uncertainties affect the values of $M_R$ and therefore also the predicted value of $\eta_B$.

\item {\it Charged fermion running masses.}
The predictive relations involve one of the running charged fermion mass matrices, $\hat m_u$, $\hat m_d$, or $\hat m_e$, depending on the model under consideration. This introduces a systematic uncertainty in $M_R$ and $m_\nu^D$, due to the choice of renormalization scale and running scheme. In addition, different EFT setups, such as SMEFT and SMEFT+$N_R$, can lead to small differences in the matching and running, which give an additional theoretical uncertainty in the input masses.

In {\bf Model A}, the strong hierarchy in the up-type quark masses leads to a strongly hierarchical $M_R$, so the result is highly sensitive to the running quark masses. In particular, the top quark plays an important role. The value of the top Yukawa coupling depends on how the top mass is measured experimentally and on how one converts the top pole mass into the $\overline{\rm MS}$ running mass. This conversion involves QCD and electroweak corrections. As a result, even a small change in the top mass input can lead to a noticeable change in the reconstructed $M_R$ and in the leptogenesis washout parameters.

In {\bf Model B}, the uncertainty comes from the down-type quark masses. These masses, especially for the light quarks, are not directly measured and are affected by low energy QCD effects. Therefore, when they are evolved to high scales, the resulting running masses depend on the choice of renormalization scale, matching procedure, and the hadronic inputs used in their extraction.

In {\bf Model C}, the charged lepton masses are known very precisely from experiment, so the corresponding uncertainty is usually smaller. It mainly comes from the choice of scale used in the matching and running.

\item {\it Spectator and thermal corrections.} Compared with the case where spectator effects are neglected, including them can have a sizable impact on the final baryon asymmetry. They may also change which lepton flavor gives the dominant contribution to the final lepton asymmetry. In addition, different ways of treating spectator effects can lead to some further systematic uncertainty, as we will discuss in Sec.~\ref{subsec:finite-spectator}.

\end{itemize}

Moreover, we neglect certain processes that will not give a dominant contribution to our analysis, such as $\Delta L=2$ washout and $\Delta L=1$ scatterings, etc.~\cite{Buchmuller:2004nz}. The effects discussed above, along with other next-to-leading order corrections, lead to an uncertainty of $\mathcal{O}(1)$ in the final baryon asymmetry~\cite{Salvio:2011sf,Garbrecht:2019zaa}. In the numerical scan, the predicted $\eta_B$ therefore need not lie within a very narrow interval around the observed value. Instead, we use a broader phenomenologically acceptable range, since it reveals the relevant regions of parameter space more clearly. For the benchmark points, representative examples are selected, with preference given to those that predict a baryon asymmetry closest to the observed value.

\subsection{Numerical Results}
\label{subsec:results}
This section presents the main numerical results of the analysis. It begins with the scan results, which identify the viable regions of parameter space and their characteristic heavy neutrino mass patterns. Based on these results, representative benchmark points are then selected for a more detailed study, including the impact of finite rate spectator effects. Finally, the implications of the scan results for neutrinoless double beta decay are discussed.

\subsubsection{Scanning Results}

The purpose of the numerical scan is to identify the regions of parameter space that can reproduce a phenomenologically acceptable baryon asymmetry, and to determine the associated heavy neutrino mass patterns and correlations with low energy parameters. The scan is performed over four parameters: the lightest neutrino mass $m_{\rm lightest}\in[10^{-4},10^{-1}]~{\rm eV}$, Dirac CP phase $\delta_{\rm CP}\in[0,2\pi]$, and Majorana phases $\alpha,\beta\in[0,\pi]$. As discussed above, a relatively broad acceptance range, $\eta_{B}\in[5,7]\times10^{-10}$, is used in the scan to account for the $\mathcal{O}(1)$ theoretical uncertainty. 

To improve the efficiency of the parameter scan, a two-step strategy is adopted. The first step consists of a broad scan over about $3\times10^7$ random points for each model. This scan estimates the typical size of the generated baryon asymmetry and locates potentially viable regions of parameter space. Points with $\eta_B$ close to the observed value appear only for normal ordering (NO) of light neutrino masses, while no viable region is found for inverted ordering (IO). The following discussion therefore concentrates on the NO case. The scan also shows that viable points do not spread over the full parameter space, but stay in regions where two heavy neutrino masses are close.

In the second step, the close-mass points found in the broad scan were used as seeds for a local refinement. In Model B, this seed sample consists of 341 points in the $M_1\simeq M_2$ pair and 500 points in the $M_2\simeq M_3$ pair. In Model C, the seed sample contains 1032 points, all belonging to the $M_2\simeq M_3$ pair. Around each seed point, 20 rounds of local refinement were carried out, with $2\times10^4$ points generated in each round. For the refined samples, only points satisfying $|M_i-M_j|/M_i<10^{-3}$ for the relevant close-mass pair were kept, and the baryon asymmetry $\eta_B$ was then recomputed using the full density matrix evolution. 

\paragraph{Model A} For the up-type quark Dirac spectrum, the explored parameter region does not yield the observed baryon asymmetry. Even after refining the scan around an $N_1$-$N_2$ close-mass region with $(M_2-M_1)/M_1<10^{-4}$, the largest asymmetry remains only of order $10^{-11}$. This indicates that, in Model A, the CP asymmetry remains too small, while the corresponding washout effects are still sizable. Model A is therefore not considered further in the benchmark analysis, and the discussion below focuses on Models B and C.

\begin{figure}[H]
    \centering
    \includegraphics[width=0.7\textwidth]{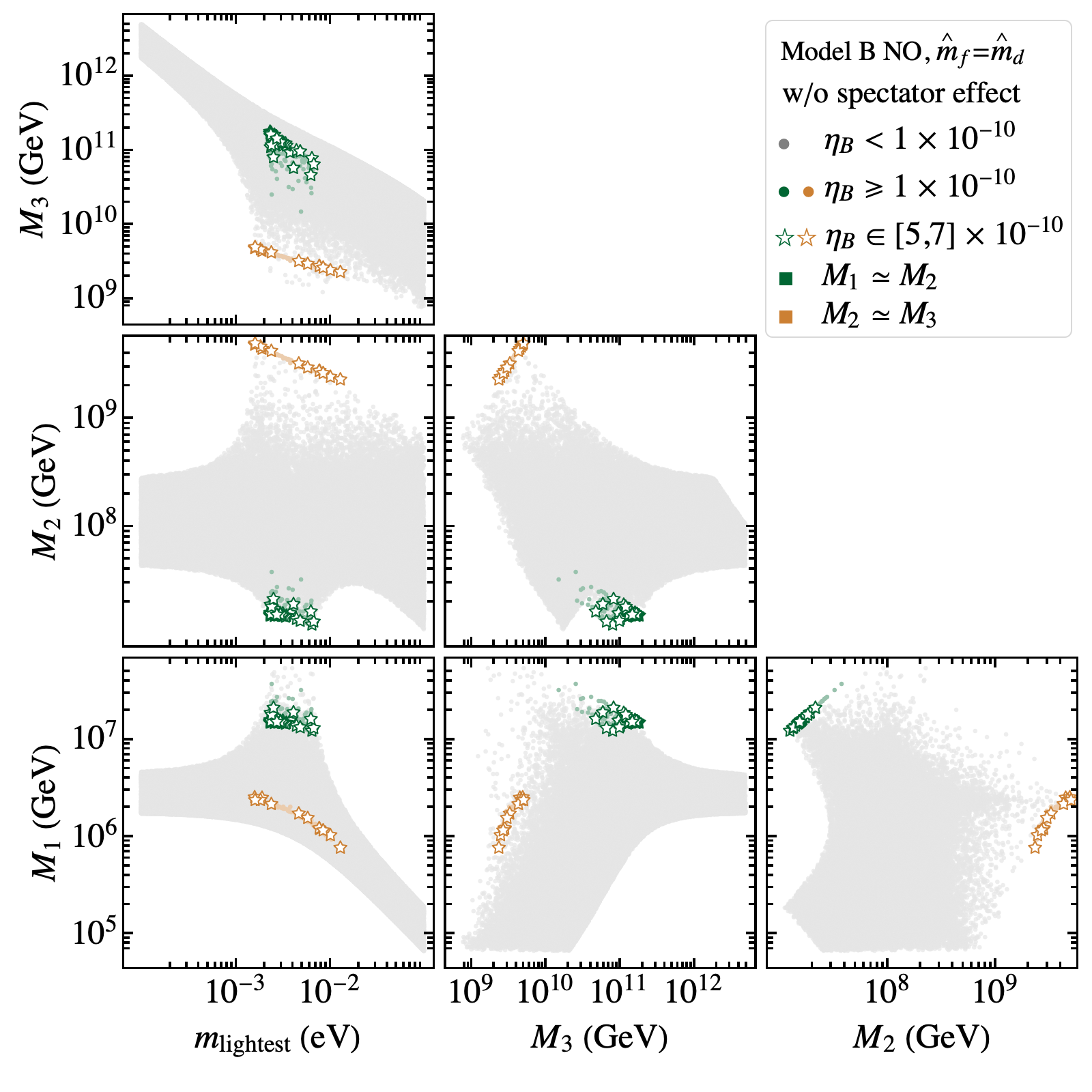}
    \includegraphics[width=0.7\textwidth]{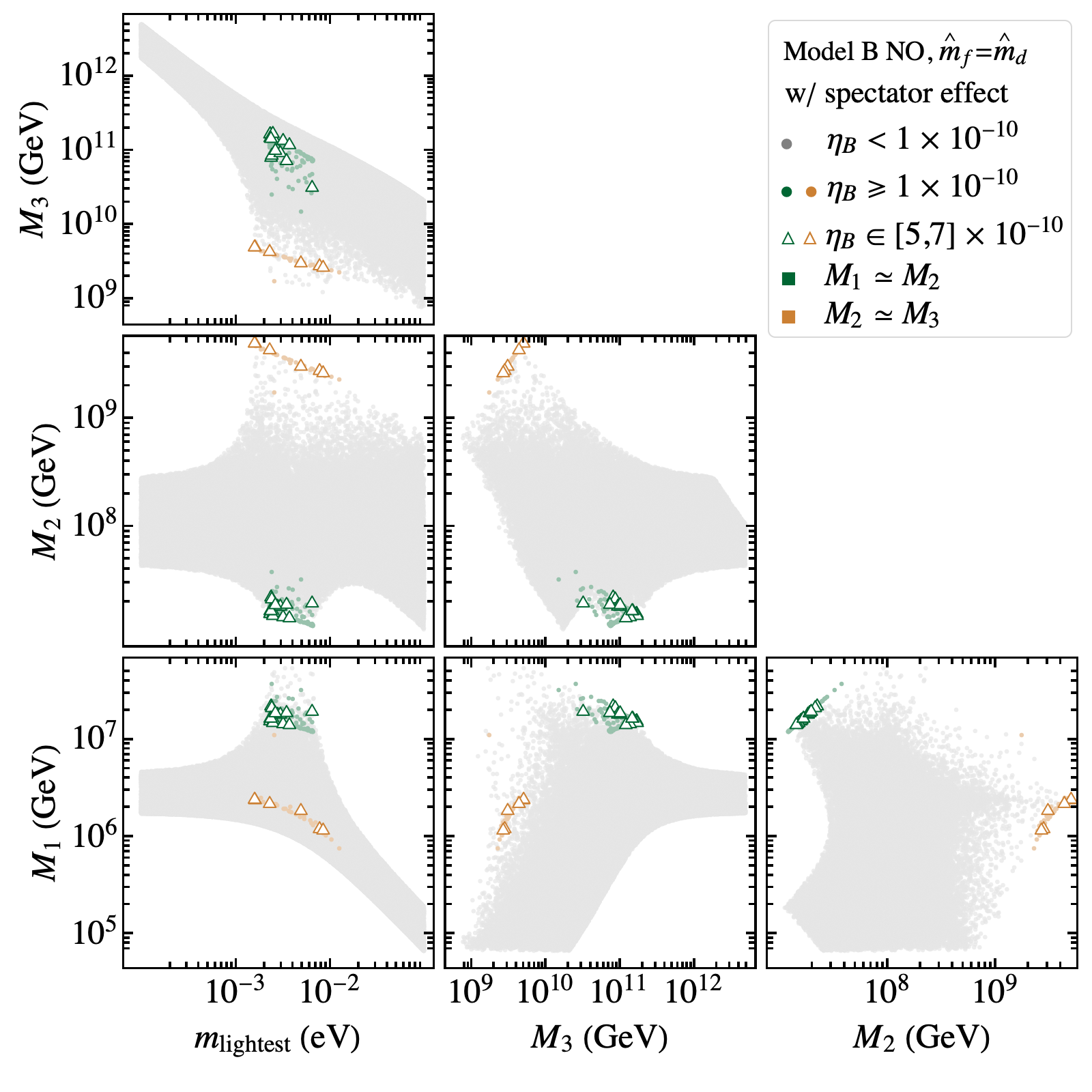}
    \caption{Scanning results in model B with the light neutrino masses in normal ordering (NO), showing the relations between the lightest neutrino mass $m_{\rm lightest}$ and the heavy RH neutrino masses $M_{i}$. For more details, see the main text.}
    \label{B-scan-1}
\end{figure}

\begin{figure}[H]
    \centering
    \includegraphics[width=0.7\textwidth]{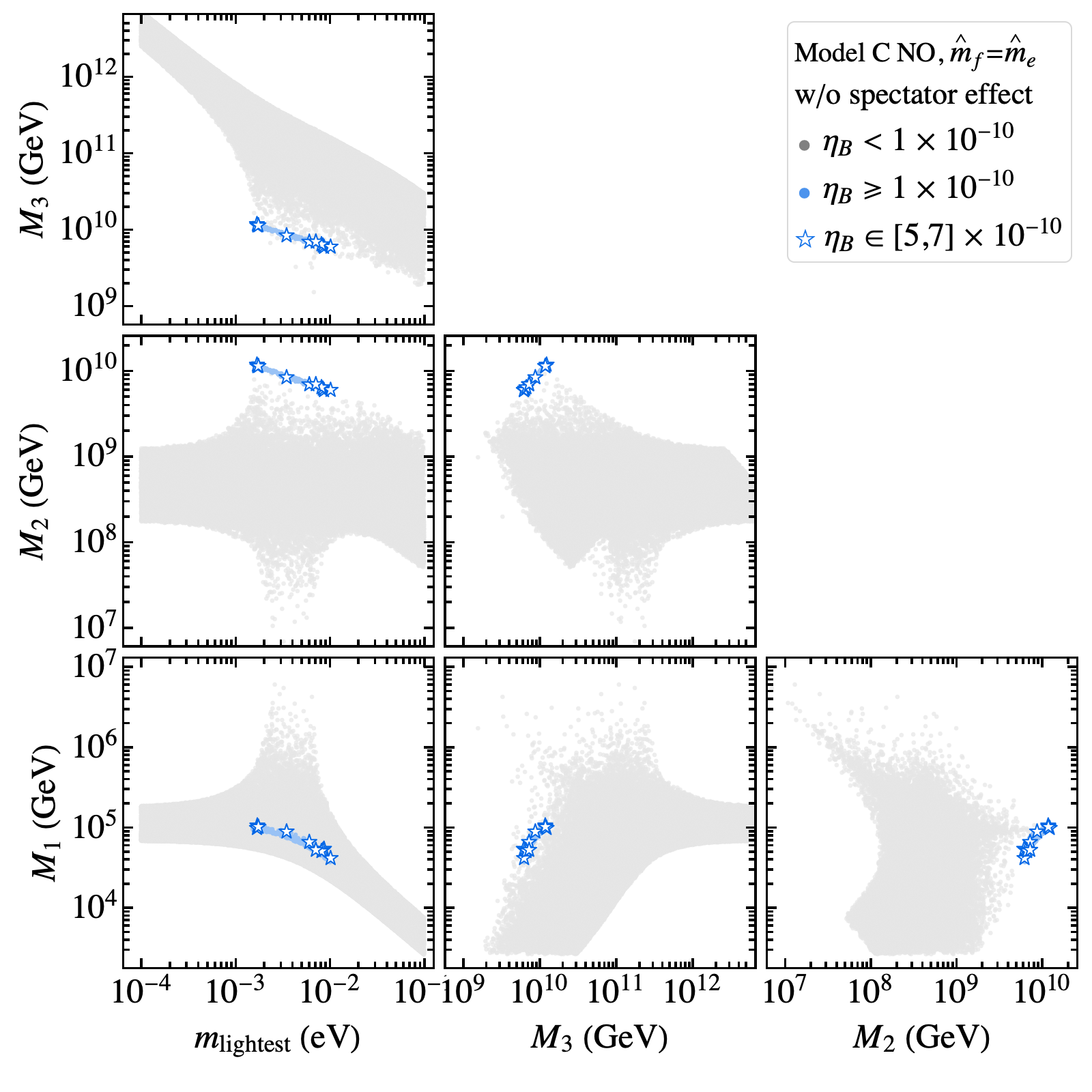}
    \includegraphics[width=0.7\textwidth]{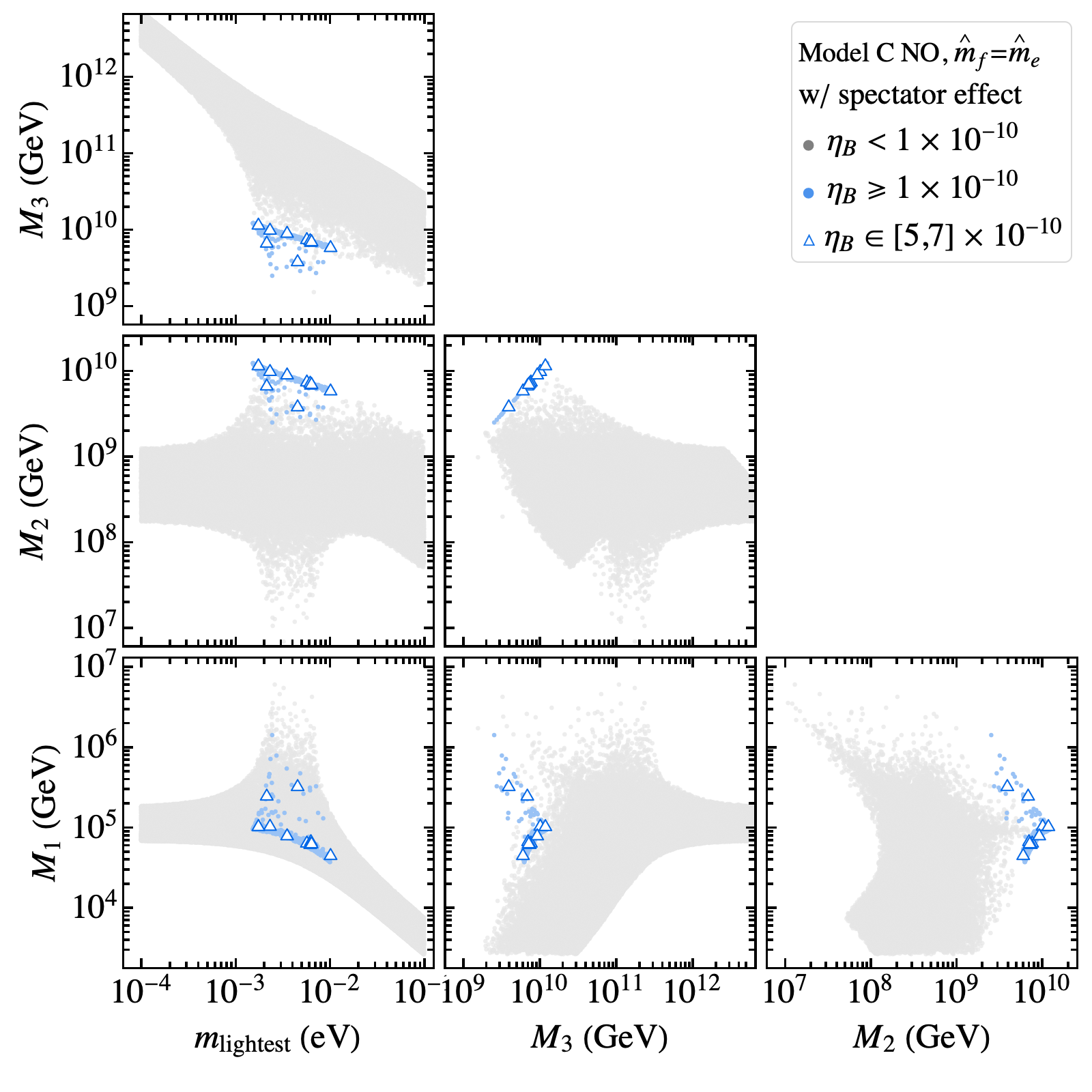}
    \caption{Scanning results in model C with the light neutrino masses in normal ordering (NO), showing the relations between the lightest neutrino mass $m_{\rm lightest}$ and the heavy RH neutrino masses $M_{i}$. For more details, see the main text.}
    \label{C-scan-1}
\end{figure}

\begin{figure}[H]
    \centering
    \includegraphics[width=0.7\textwidth]{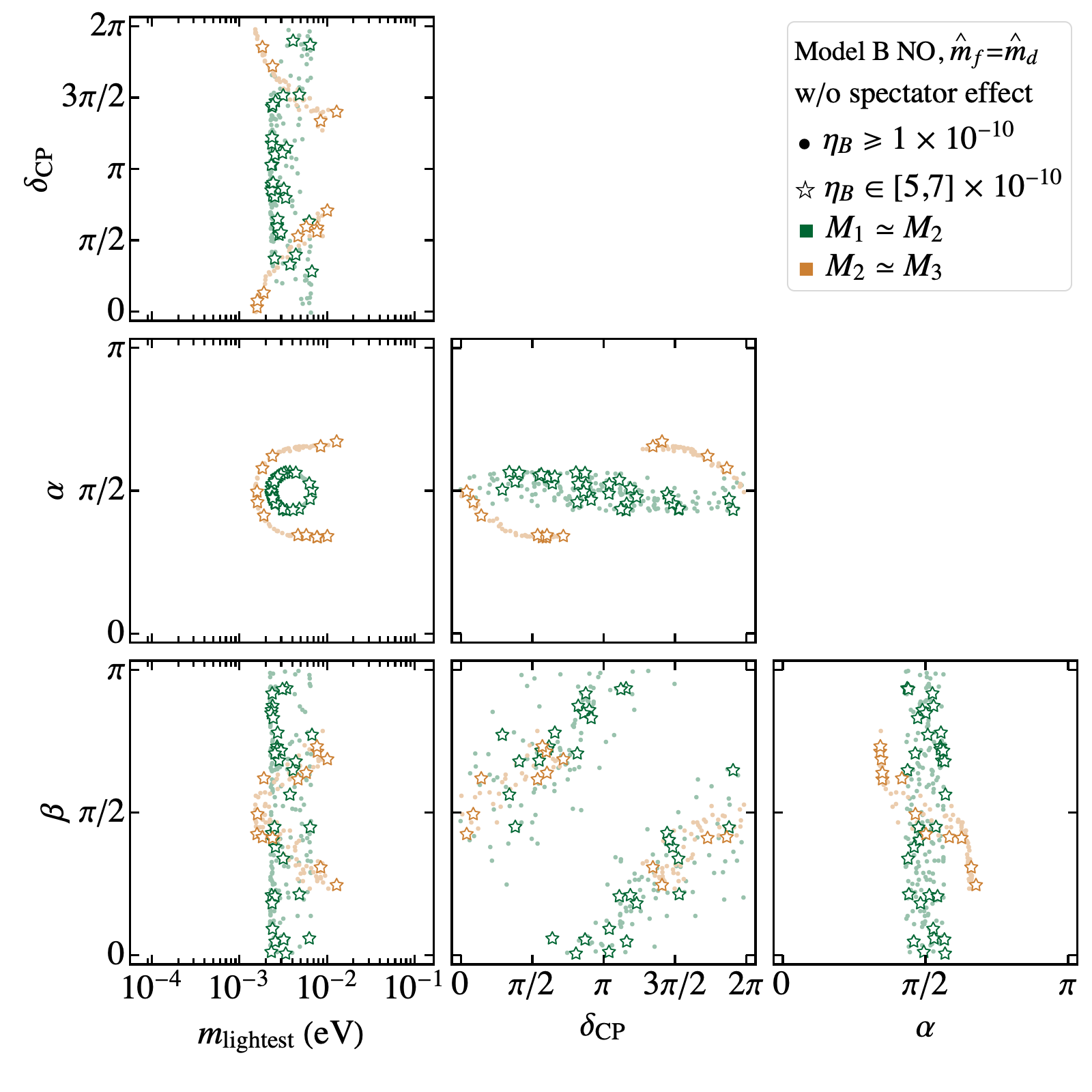}\\
    \includegraphics[width=0.7\textwidth]{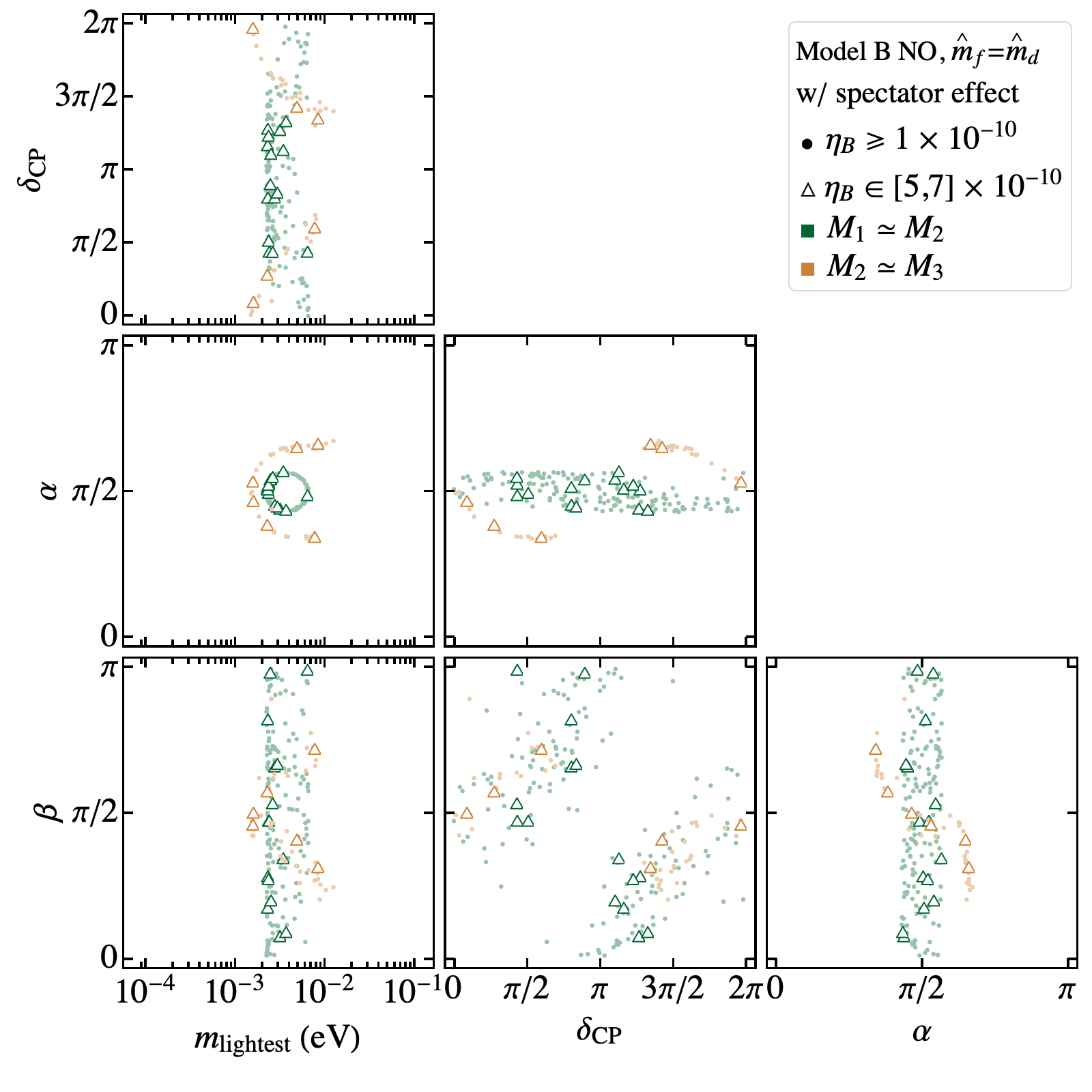}
    \caption{Scanning results in model B with the light neutrino masses in normal ordering (NO), showing the correlation between $m_{\rm lightest}$ and phases $\delta_{\rm CP}$, $\alpha$ and $\beta$. For more details, see the main text.}
    \label{B-scan-2}
\end{figure}

\begin{figure}[H]
    \centering
    \includegraphics[width=0.7\textwidth]{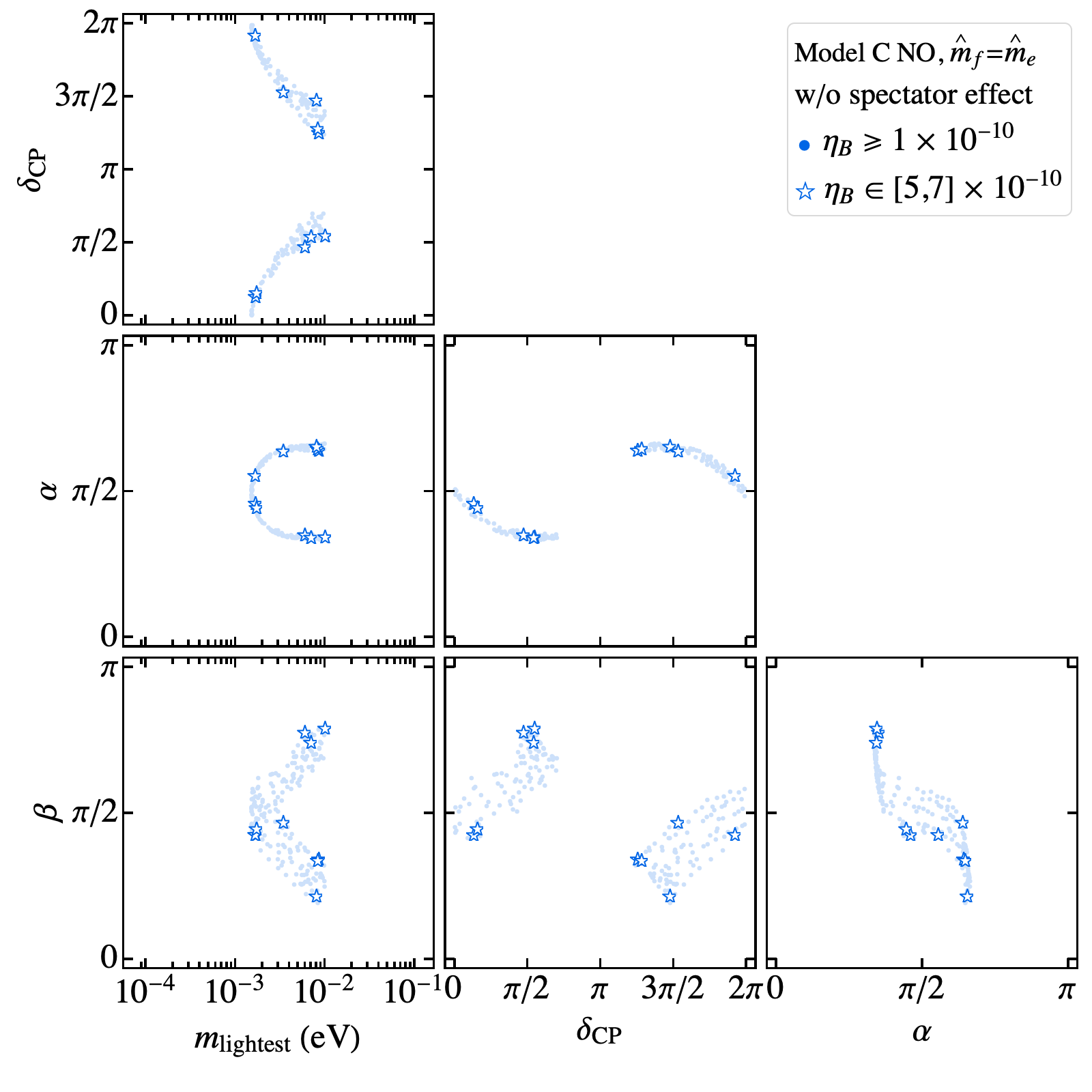}
    \includegraphics[width=0.7\textwidth]{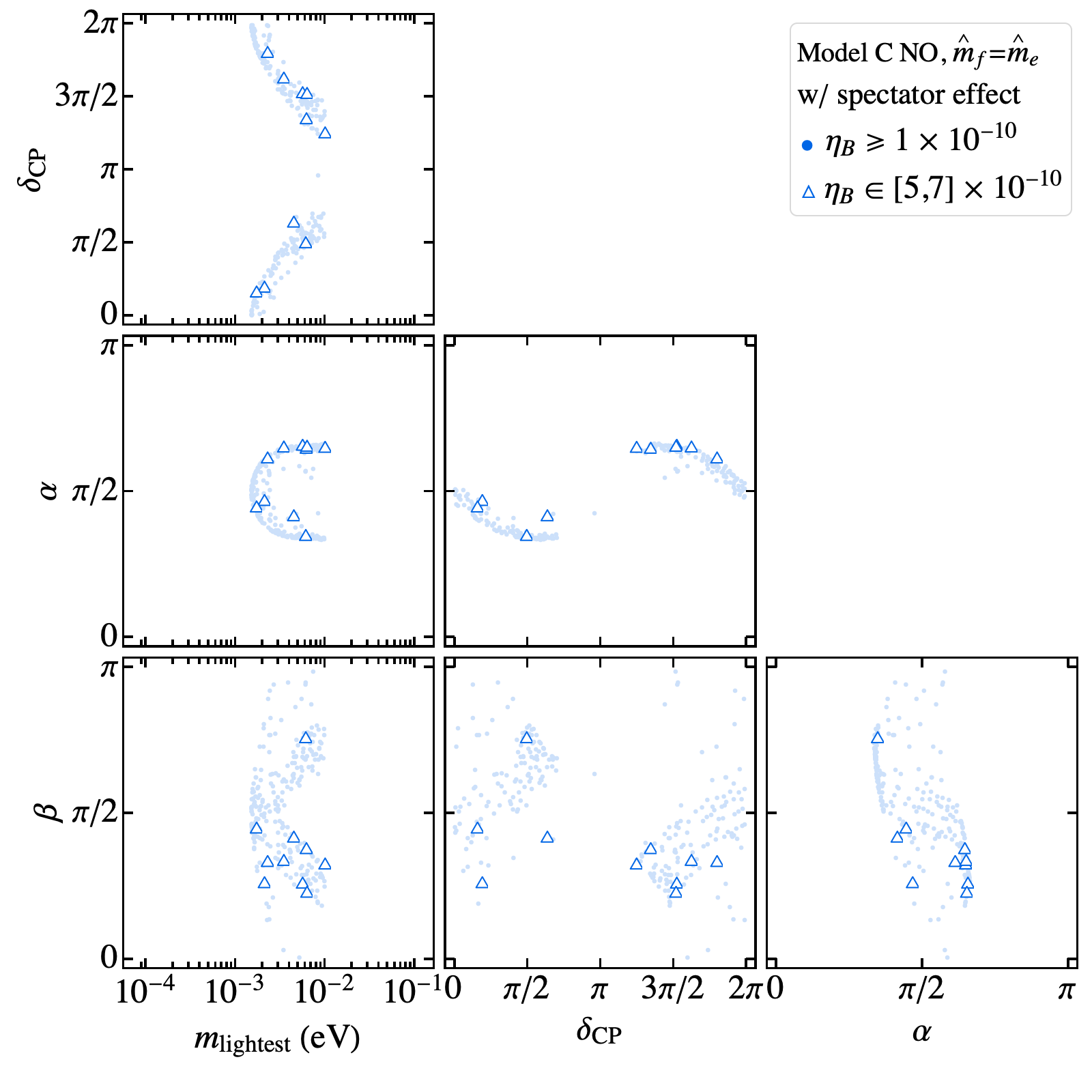}
    \caption{Scanning results in model C with the light neutrino masses in normal ordering (NO), showing the correlation between $m_{\rm lightest}$ and phases $\delta_{\rm CP}$, $\alpha$ and $\beta$. For more details, see the main text.}
    \label{C-scan-2}
\end{figure}

\paragraph{Models B and C} The main scan results for Models B and C with normal ordering (NO) are shown in Figs.~\ref{B-scan-1}-\ref{C-scan-2}, respectively. In each figure, the upper panel shows the evolution without spectator effects, while the lower panel includes the instantaneous $C^{(2)}$-$C^{(3)}$ spectator switch. Figures~\ref{B-scan-1} and \ref{C-scan-1} show the reconstructed heavy neutrino masses $M_{i}$ in GeVs together with the lightest neutrino mass $m_{\rm lightest}$, while Figs.~\ref{B-scan-2} and \ref{C-scan-2} show the correlations with the low energy phases $\delta_{\rm CP}$, $\alpha$, and $\beta$. In the figures, grey points denote $\eta_B<10^{-10}$, colored points denote $\eta_B\geq10^{-10}$, and open markers highlight the points in the acceptable range $\eta_B\in[5,7]\times10^{-10}$.

Successful leptogenesis is found only in a limited part of the parameter space, especially for the range $m_{\rm lightest}\in[10^{-3},10^{-2}]~\rm eV$. In both Models B and C, points with $\eta_B$ close to the observed value appear only when two heavy neutrino masses are close to each other. Model B, as shown in Figs.~\ref{B-scan-1}, allows viable points in both the $M_1\simeq M_2$ (green points) and $M_2\simeq M_3$ (orange points) close-mass regions. Model C, as shown in Figs.~\ref{C-scan-1}, favors mainly the $M_2\simeq M_3$ (blue points) region. Including spectator effects does not change this overall picture. The same close-mass regions remain viable, although the final value of $\eta_B$ and the density of successful points are shifted. Note that, although the viable regions involve two close heavy neutrino masses, this does not imply resonant leptogenesis. The reason why these successful points remain distinct from the resonant regime will be discussed in detail in the next subsection.

The Figs.~\ref{B-scan-2} and \ref{C-scan-2} show successful points confined to restricted regions of the low energy parameter space rather than filling it evenly. They also indicate that the Dirac phase $\delta_{\rm CP}$ alone does not suffice to generate the observed baryon asymmetry, with successful points appearing only for nonzero Majorana phases. This points to an essential role for both $\alpha$ and $\beta$. Another common feature of Models B and C concerns the viable points in the $M_2\simeq M_3$ pair, which occupy very similar regions of phase space, with $\alpha$ concentrated around $\pi/2$, $\delta_{\rm CP}$ avoiding the interval $[3\pi/4,\,5\pi/4]$, and $\beta$ lying mostly in the range $[\pi/4,\,3\pi/4]$. By contrast, the $M_1\simeq M_2$ pair in Model B shows $\alpha$ even more tightly concentrated around $\pi/2$, together with much broader allowed ranges for $\delta_{\rm CP}$ and $\beta$.

\vskip 0.4cm
These features motivate the benchmark selection discussed in the next subsection. The evolution of the lepton and baryon asymmetries for representative benchmark points in Models B and C will be examined in detail, in order to show how the observed baryon asymmetry is generated, and how the different treatment of spectator effects influences the final result. Particular attention will be paid to the origin of the enhancement in the close-mass region and to its distinction from the usual resonant leptogenesis mechanism.

\subsubsection{Benchmark Points}
According to the scan results discussed above, two benchmark points (BPs) are chosen for each of Models B and C from the successful scan points that give values of $\eta_B$ closest to the experimental result.
In Model B, the two points are taken from the parameter regions with $M_1 \simeq M_2$ and $M_2 \simeq M_3$, respectively. In Model C, the two points are chosen such that their parameter values show a relatively large separation.
\begin{table}[t]
    \centering
    \resizebox{0.8\textwidth}{!}{
    \renewcommand{\arraystretch}{0.95}
    \begin{tabular}{c|c|c|c|c}
      \hline
        Para. & B, BP1 & B, BP2 & C, BP1 & C, BP2 \\ \hline\hline
        \rowcolor{blockgray}$m_{\rm{lightest}}$ & 0.001557  & 0.002369  & 0.005583  & 0.009959  \\ 
        \rowcolor{blockgray}$\delta_{\rm{CP}}$ & 6.15933  & 1.34165  & 4.77529  & 3.91064  \\ 
        \rowcolor{blockgray}$\alpha$ & 1.66069  & 1.63764  & 2.05725  & 2.03482  \\
        \rowcolor{blockgray}$\beta$ & 1.43194  & 1.47476  & 0.80768  & 1.01377  \\
        \hline
        \rowcolor{blockblue}$M_1$ & 2.4181127$\times 10^{6}$ & 2.1204273$\times 10^{7}$ & 6.4710113$\times 10^{4}$ & 4.5254428$\times 10^{4}$ \\ 
        \rowcolor{blockblue}$M_2$ & 4.9651357$\times 10^{9}$ & 2.1204309$\times 10^{7}$ & 7.5011335$\times 10^{9}$ & 5.8975143$\times 10^{9}$ \\ 
        \rowcolor{blockblue}$M_3$ & 4.9655608$\times 10^{9}$ & 8.5369030$\times 10^{10}$ & 7.5012545$\times 10^{9}$ & 5.8976828$\times 10^{9}$ \\ \hline
        \rowcolor{blockgreen}$\epsilon_{1e}$ & -2.762$\times 10^{-14}$ & -9.220$\times 10^{-8}$ & -2.096$\times 10^{-17}$ & 1.369$\times 10^{-17}$ \\ 
        \rowcolor{blockgreen}$\epsilon_{1\mu}$ & 1.102$\times 10^{-11}$ & 3.683$\times 10^{-5}$ & 9.340$\times 10^{-13}$ & -6.110$\times 10^{-13}$ \\ 
        \rowcolor{blockgreen}$\epsilon_{1\tau}$ & 2.022$\times 10^{-11}$ & 4.192$\times 10^{-5}$ & 1.060$\times 10^{-12}$ & -4.198$\times 10^{-13}$ \\ 
        \rowcolor{blockgreenhi}$\epsilon_{1\text{tot}}$ & 3.121$\times 10^{-11}$ & 7.865$\times 10^{-5}$ & 1.994$\times 10^{-12}$ & -1.031$\times 10^{-12}$ \\ \hline
        \rowcolor{blockgreen}$\epsilon_{2e}$ & -4.563$\times 10^{-11}$ & -1.272$\times 10^{-7}$ & -7.785$\times 10^{-15}$ & -1.940$\times 10^{-14}$ \\ 
        \rowcolor{blockgreen}$\epsilon_{2\mu}$ & -7.932$\times 10^{-7}$ & 5.081$\times 10^{-5}$ & -3.338$\times 10^{-6}$ & -1.163$\times 10^{-5}$ \\ 
        \rowcolor{blockgreen}$\epsilon_{2\tau}$ & 6.752$\times 10^{-4}$ & 5.783$\times 10^{-5}$ & 8.940$\times 10^{-4}$ & 3.113$\times 10^{-3}$ \\ 
        \rowcolor{blockgreenhi}$\epsilon_{2\text{tot}}$ & 6.744$\times 10^{-4}$ & 1.085$\times 10^{-4}$ & 8.907$\times 10^{-4}$ & 3.102$\times 10^{-3}$ \\ \hline
        \rowcolor{blockgreen}$\epsilon_{3e}$ & -4.132$\times 10^{-11}$ & 8.115$\times 10^{-18}$ & -6.881$\times 10^{-15}$ & -1.824$\times 10^{-14}$ \\ 
        \rowcolor{blockgreen}$\epsilon_{3\mu}$ & -7.183$\times 10^{-7}$ & 8.105$\times 10^{-17}$ & -2.951$\times 10^{-6}$ & -1.093$\times 10^{-5}$ \\ 
        \rowcolor{blockgreen}$\epsilon_{3\tau}$ & 6.114$\times 10^{-4}$ & -2.768$\times 10^{-12}$ & 7.903$\times 10^{-4}$ & 2.927$\times 10^{-3}$ \\ 
        \rowcolor{blockgreenhi}$\epsilon_{3\text{tot}}$ & 6.107$\times 10^{-4}$ & -2.768$\times 10^{-12}$ & 7.873$\times 10^{-4}$ & 2.916$\times 10^{-3}$ \\ \hline
        \rowcolor{blockyellow}$K_{1e}$ & 2.530  & 0.1303  & 3.869  & 5.532  \\ 
        \rowcolor{blockyellow}$K_{1\mu}$ & 23.26  & 65.84  & 18.06  & 18.49  \\
        \rowcolor{blockyellow}$K_{1\tau}$ & 2.099  & 82.27  & 2.588  & 3.663  \\ 
        \rowcolor{blockyellowhi}$K_{1\text{tot}}$ & 27.89  & 148.2  & 24.51  & 27.68  \\ \hline
        \rowcolor{blockyellow}$K_{2e}$ & 1.475$\times 10^{-5}$ & 0.1648  & 1.832$\times 10^{-9}$ & 1.559$\times 10^{-9}$ \\
        \rowcolor{blockyellow}$K_{2\mu}$ & 0.2587  & 52.07  & 0.7931  & 0.9793  \\ 
        \rowcolor{blockyellow}$K_{2\tau}$ & 199.4  & 55.21  & 187.6  & 246.5  \\ 
        \rowcolor{blockyellowhi}$K_{2\text{tot}}$ & 199.6  & 107.4  & 188.4  & 247.4  \\ \hline
        \rowcolor{blockyellow}$K_{3e}$ & 1.358$\times 10^{-5}$ & 3.972$\times 10^{-9}$ & 1.651$\times 10^{-9}$ & 1.614$\times 10^{-9}$ \\ 
        \rowcolor{blockyellow}$K_{3\mu}$ & 0.2342  & 3.945$\times 10^{-5}$ & 0.7004  & 0.9203  \\ 
        \rowcolor{blockyellow}$K_{3\tau}$ & 220.2  & 24.37  & 212.4  & 262.2  \\
        \rowcolor{blockyellowhi}$K_{3\text{tot}}$ & 220.4  & 24.37  & 213.1  & 263.1 \\ \hline
        \rowcolor{blockpink}$R_{\rm reson}^i$ & 61  & 380  & 8  & 14 \\ 
        \rowcolor{blockpink}$R_{\rm reson}^j$ & 55  & 524  & 7  & 13 \\ \hline
        \rowcolor{blockpurple}$\eta_{B}^{\rm inst}$ & 5.8779$\times 10^{-10}$ & 6.1009$\times 10^{-10}$ & 6.0628$\times 10^{-10}$ & 5.9482$\times 10^{-10}$ \\
        \rowcolor{blockpurple}$\eta_{B}^{\rm finite}$ & 6.0760$\times 10^{-10}$ & 6.1032$\times 10^{-10}$ & 6.6893$\times 10^{-10}$ & 6.5820$\times 10^{-10}$ \\ 
        \rowcolor{blockpurplehi}$\Delta\eta_{B}$ & $+3.37\%$ & $+0.04\%$ & $+10.33\%$ & $+10.66\%$ \\ \hline\hline
    \end{tabular}}
    \caption{Four benchmark points that yield an acceptable baryon asymmetry. Here, $M_i$ denotes the heavy RH neutrino masses in GeV, $m_{\rm lightest}$ is given in eV, and the phases are given in radians, while $\epsilon_{i\alpha,\rm tot}$ and $K_{i\alpha,\rm tot}$ denote the CP asymmetries and decay parameters associated with $N_{i}$ decays, respectively. The $R_{\rm reson}^{i}$ and $R_{\rm reson}^{j}$ denote the ratios $|M_i-M_j|/\Gamma^{\rm tot}_i$ and $|M_i-M_j|/\Gamma^{\rm tot}_j$, respectively, for the close-mass pair. The $\eta_B^{\rm inst}$ and $\eta_B^{\rm finite}$ denote the baryon asymmetry predicted with the instantaneous and finite-rate spectator treatment, respectively. The $\Delta\eta_{B}$ is the difference between these two treatments.}
    \label{FourBPs}
\end{table}

Table~\ref{FourBPs} collects these BPs. It contains the values of low energy free parameters $m_{\rm lightest},~\delta_{\rm CP},~\alpha$ and $\beta$ that have been input, and the corresponding heavy RH neutrino masses $M_{i}$. To aid in an explanation of how these BPs achieve the final baryon asymmetry, the CP asymmetries $\epsilon_{i\alpha}$ and $\epsilon_{i\rm tot}=\sum_{\alpha}\epsilon_{i\alpha}$, the decay parameters $K_{i\alpha}$ and $K_{i\rm tot}$ as defined in Eq.~\eqref{eq:K} are also shown. Furthermore, the values of quantities $R_{\rm reson}^{i}$ used to judge whether the close-mass pair falls into the resonant region are presented. The predicted values of $\eta_B$ from different treatments of the spectator effect, and the associated spread $\Delta \eta_{B}$, are shown in the final rows.

\paragraph{Non-Resonant Regime} To assess whether the close-mass heavy neutrino pairs enter the resonant regime, the relevant quantity $R_{\rm reson}^{i,j}$ is the ratio between the mass splitting and the decay width~\cite{Pilaftsis:2003gt,Pilaftsis:1997jf}. Using the relation and the decay width shown in Eq.~\eqref{eq:Kpara}, the ratio becomes 
\begin{align}
   R_{\rm reson}^{i,j}\equiv\dfrac{|M_{i}-M_{j}|}{\Gamma^{\rm tot}_{i,j}}
    =\dfrac{|M_{i}-M_{j}|}{M_{i,j}}
    \dfrac{4\pi v^2}{K_{i,j}m_{*}M_{i,j}}~,
\end{align}
noting again that the VEV is 246 GeV. For all BPs listed, the corresponding $R_{\rm reson}$ values are larger than 1, and even the smallest value is only about $7-8$. This indicates that the mass splitting remains well above the decay width, so these points do not lie in the conventional resonant leptogenesis regime, although the closeness of the masses can still lead to some enhancement.

\paragraph{Different Treatments of Spectator Effects}
\label{subsec:finite-spectator}

In the previous discussion, spectator effects have been described through the matrix $\mathcal{C}(z)$ in the instantaneous treatment, as implemented in Eq.~\eqref{eq:spectator-switch}. This is the standard approximation, in which a given interaction is assumed to be either fully equilibrated or completely not.

We also consider finite rate spectator effects. In this treatment, the set of relevant interactions is kept unchanged, but the muon and strange Yukawa interactions are allowed to enter equilibrium gradually, as motivated by Ref.~\cite{Garbrecht:2014kda}. The spectator conditions then change smoothly from the $\mathcal{C}^{(2)}$ limit to the $\mathcal{C}^{(3)}$ limit, rather than switching abruptly at a fixed temperature.

The difference in the final resulting baryon asymmetry between the two different treatments of spectator effects can be described as
\begin{align}
\Delta\eta_B
&\equiv
\left(\dfrac{\eta_B^{\rm finite}}
{\eta_B^{\rm inst}}-1\right)\!\times100\%~.
\label{eq:finite-spectator-shift}
\end{align}
As shown in Table~\ref{FourBPs}, the resulting shifts range from $0.04\%$ to $10.66\%$. In particular, the difference is negligible for Model B, BP2. In this case, the dominant CP asymmetry is generated by the $N_1$-$N_2$ close-mass pair at $M_{1,2}\simeq 2.12\times10^7~\mathrm{GeV}$, which lies far below the temperature range relevant for the gradual equilibration of the muon and strange Yukawa interactions, typically $T_{\mu,s}\sim10^9~\mathrm{GeV}$. By the time the asymmetry is produced and washed out, these spectator processes are already effectively in chemical equilibrium, so the instantaneous treatment provides an excellent approximation. For the other three BPs, by contrast, the relevant $M_2\simeq M_3$ pair takes place at temperatures of order $10^9~\mathrm{GeV}$, within the equilibration region of the muon and strange Yukawa interactions. In these cases, the finite rate treatment leads to a modest enhancement of the final baryon asymmetry relative to the instantaneous treatment.

\begin{figure}[h]
    \centering
    \includegraphics[height=0.215\textheight]{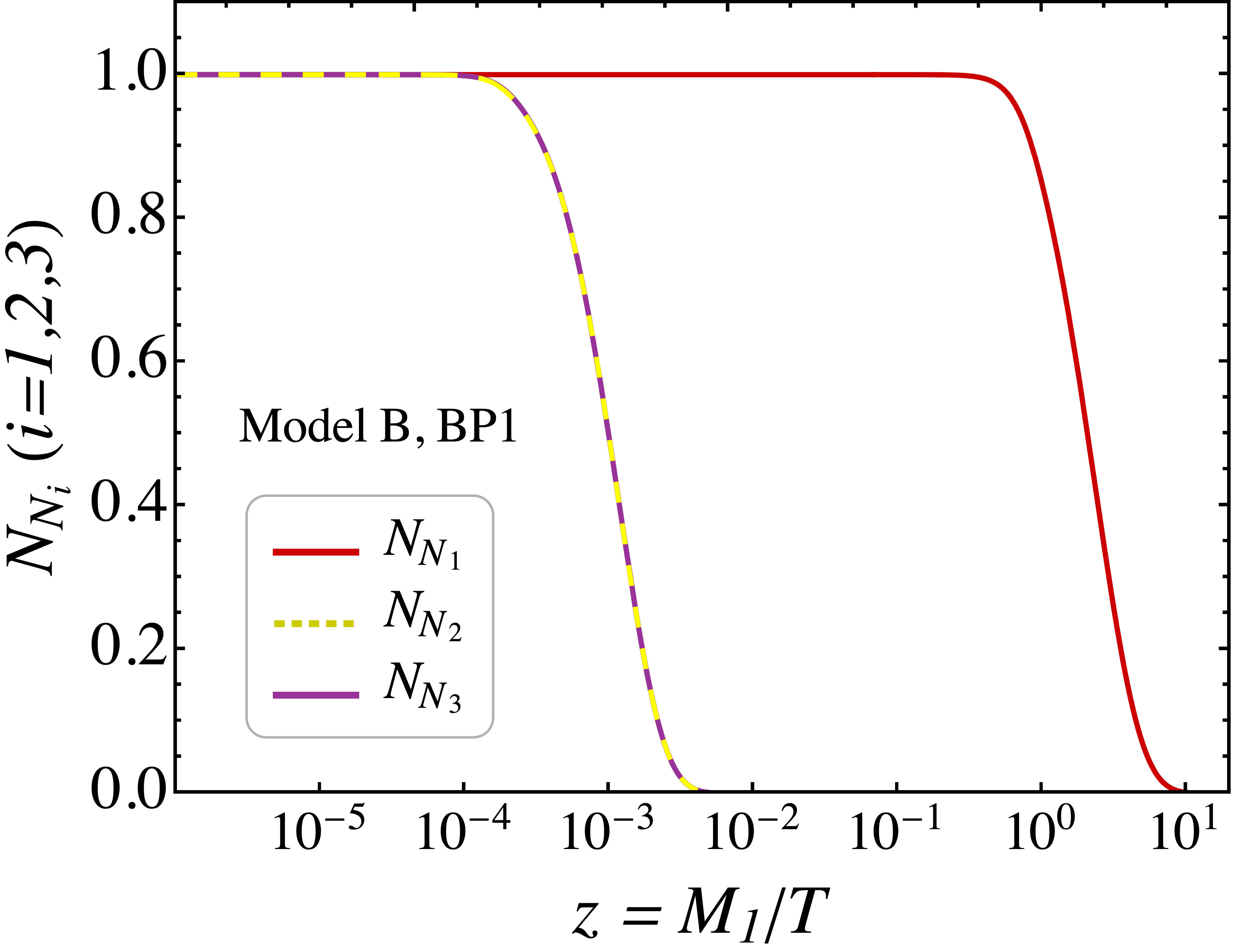}\quad
     \includegraphics[height=0.215\textheight]{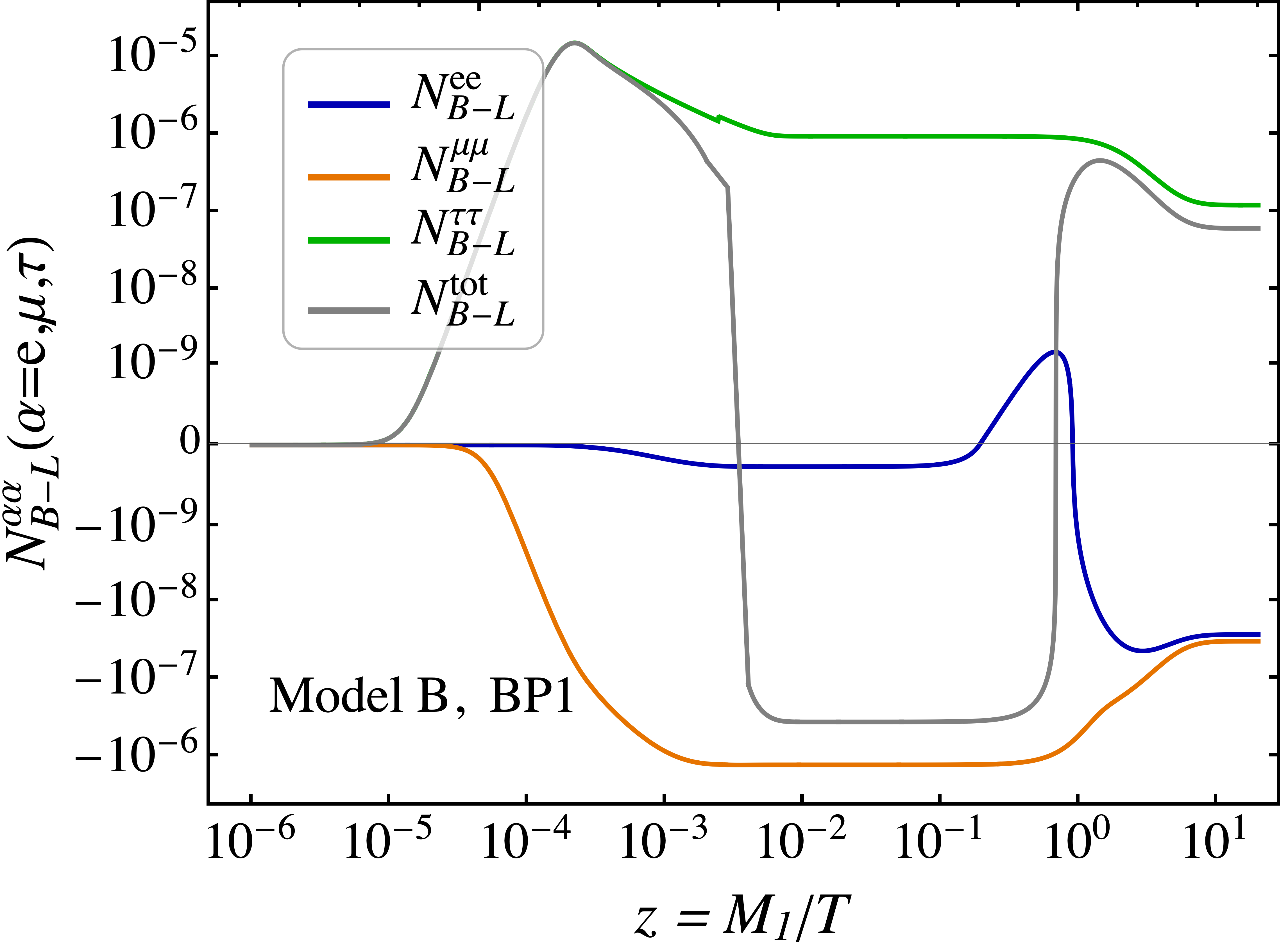}\\
    \includegraphics[height=0.215\textheight]{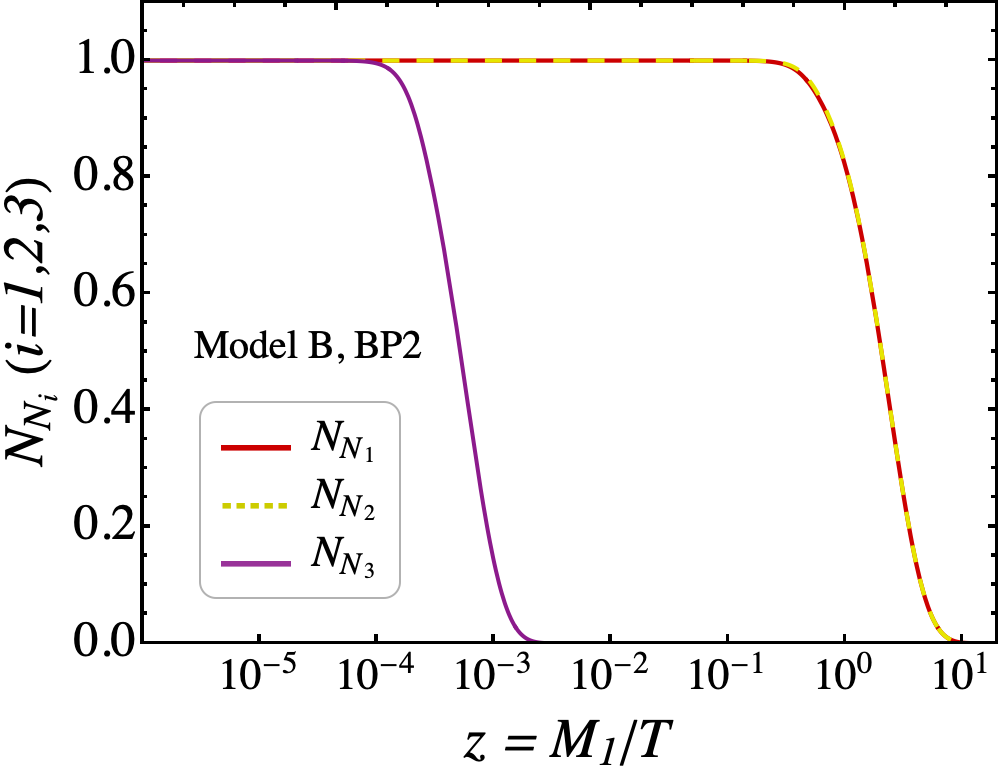}\quad
     \includegraphics[height=0.215\textheight]{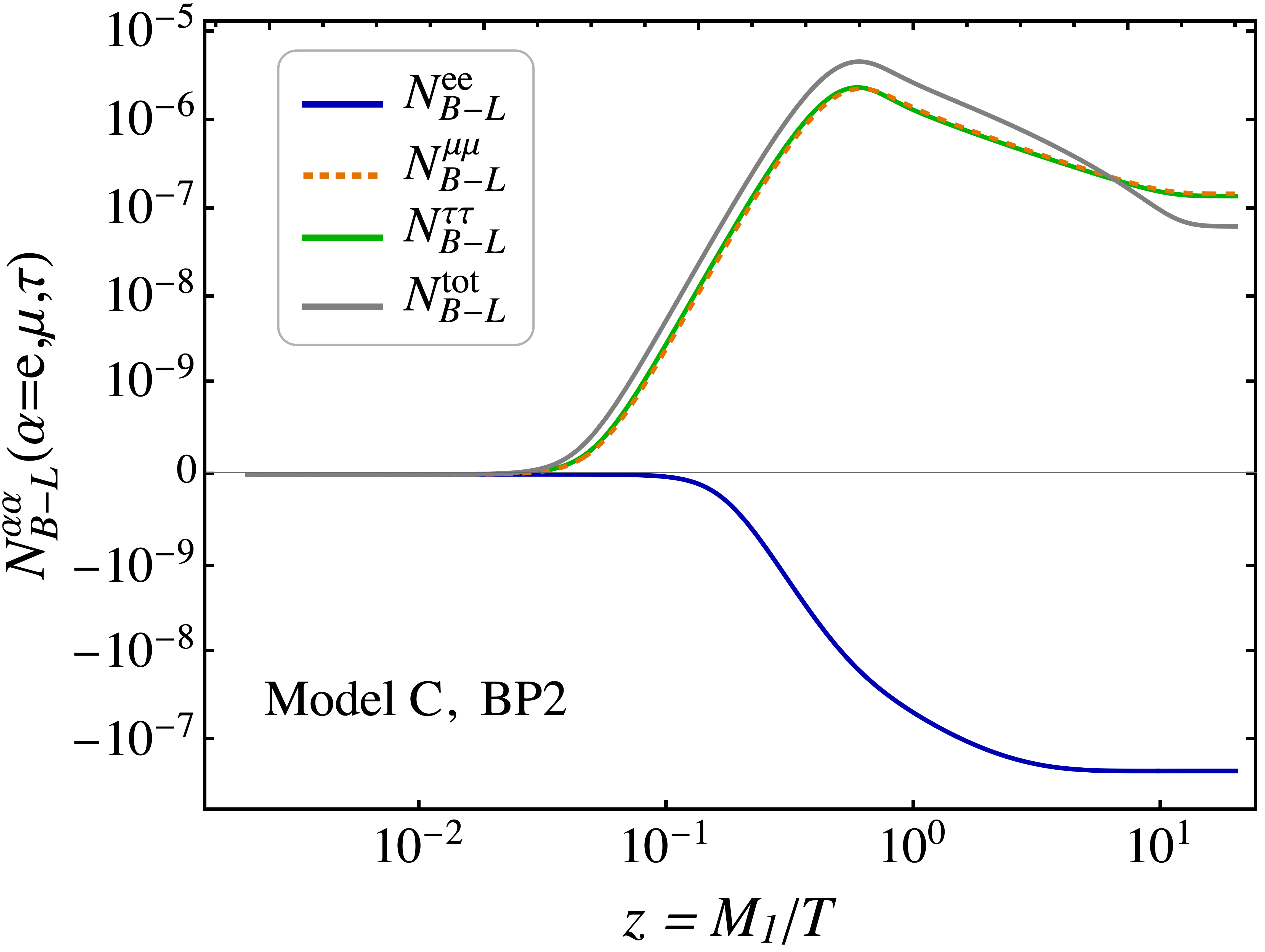}\\
    \includegraphics[height=0.215\textheight]{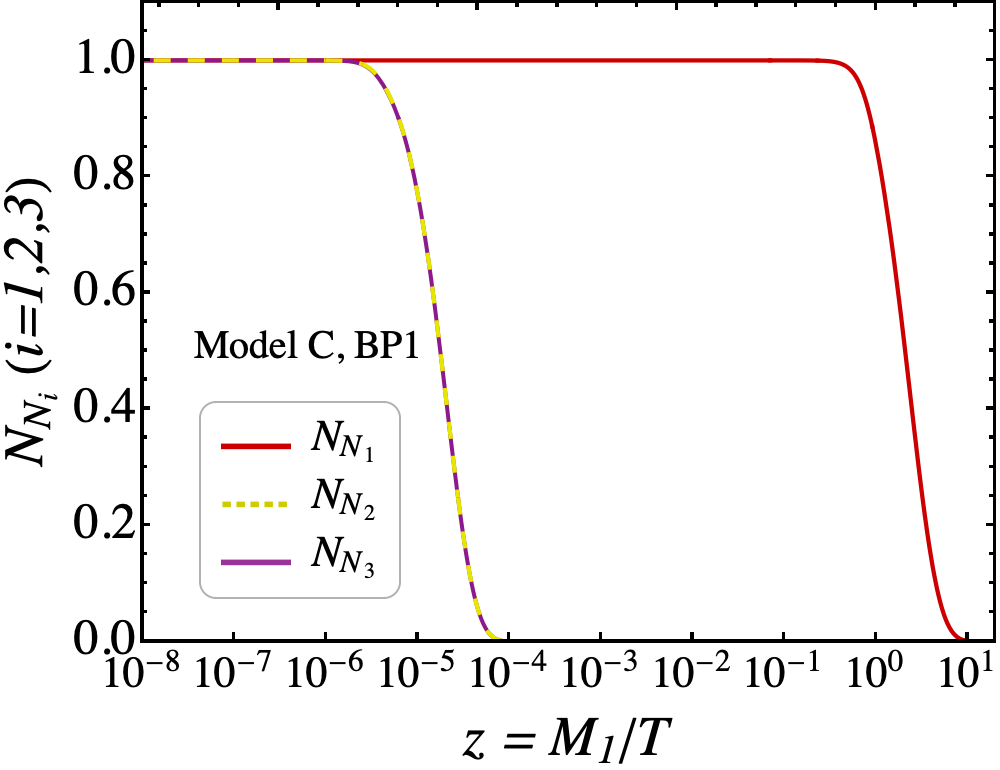}\quad
     \includegraphics[height=0.215\textheight]{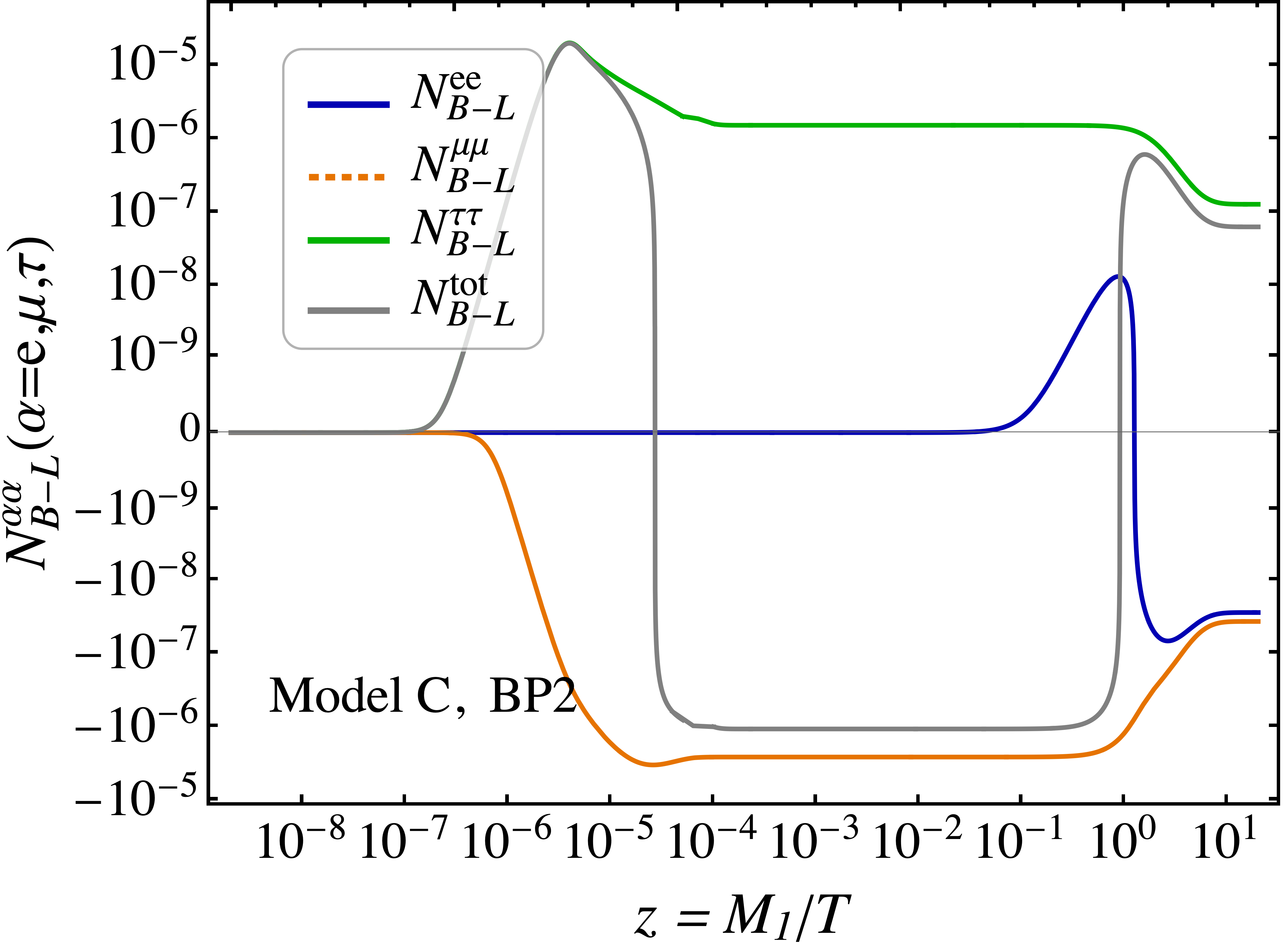}\\
    \includegraphics[height=0.215\textheight]{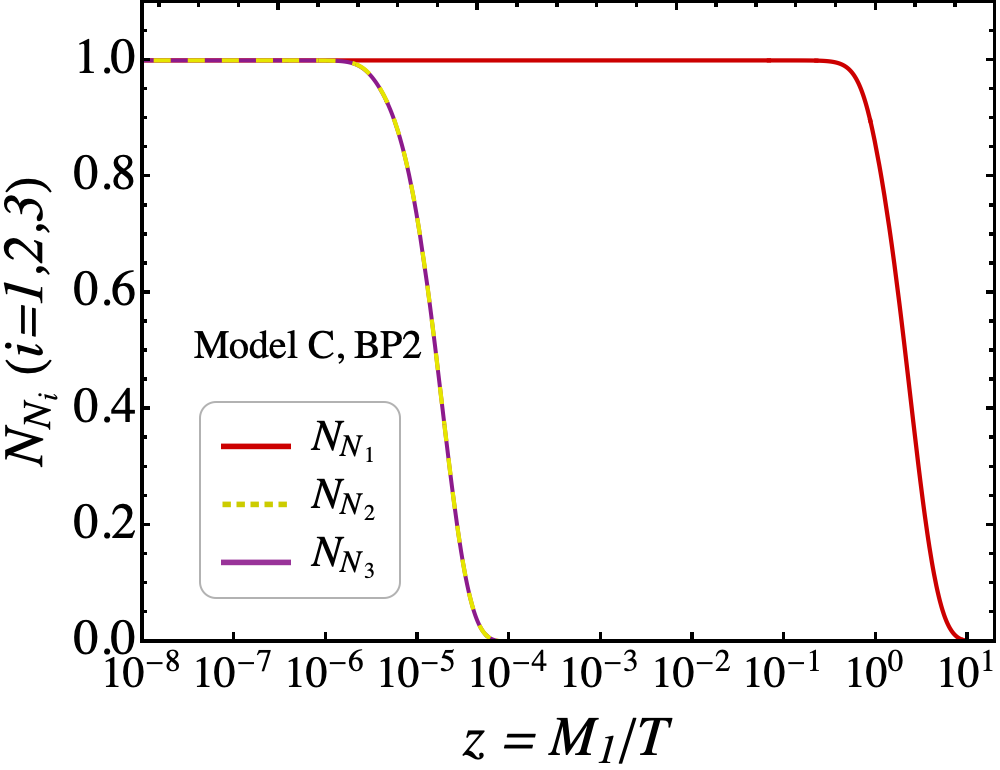}\quad
     \includegraphics[height=0.215\textheight]{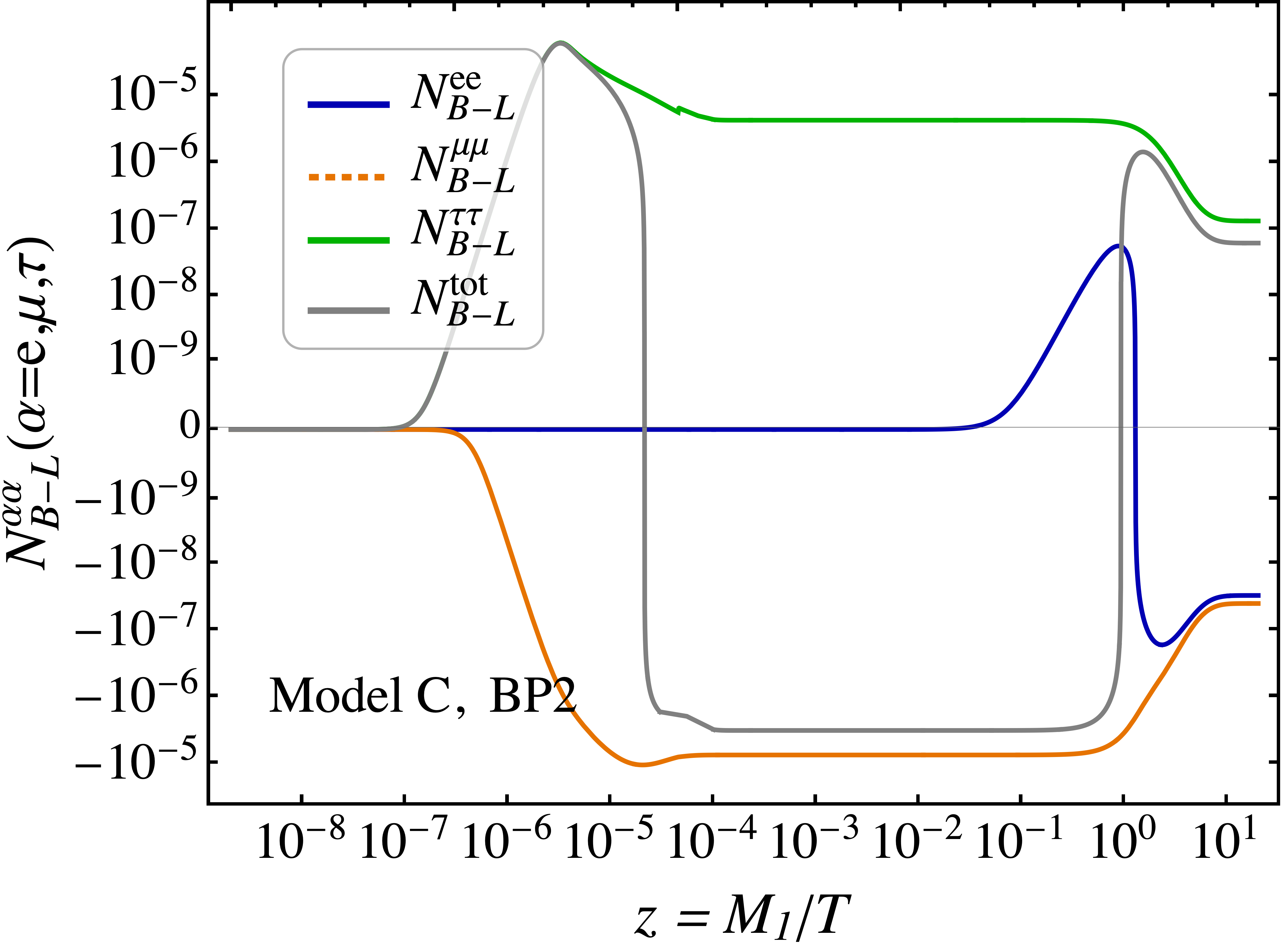}
    \caption{Evolution of the heavy RH neutrino abundances $N_{N_i}$ (left panels) and flavored $B-L$ asymmetries $N^{\alpha\alpha}_{B-L}$ together with the total asymmetry $N_{B-L}^{\rm tot}$ (right panels) for the four benchmark points.}
    \label{fig:NBL_evo}
\end{figure}

\paragraph{Evolutions of the $N_{N_i}$ and $N_{B-L}$}
Figure~\ref{fig:NBL_evo}
shows the evolution of the heavy RH neutrino $N_{i}$ abundances $N_{N_i}$ and flavored $B-L$ asymmetries $N^{\alpha\alpha}_{B-L}~(\alpha=e,\mu,\tau)$ with the variable $z=M_{1}/T$, which have been defined in Eq.~\eqref{eq:BE-RHN} and \eqref{eq:BE-density}. The total $B-L$ asymmetry is obtained by $N_{B-L}^{\rm tot}={\rm{Tr}}[N_{B-L}]=\sum_{\alpha}N^{\alpha\alpha}_{B-L}$, leading to the final $B-L$ asymmetry after the evolution given by $N_{B-L}^{\rm f}=N_{B-L}^{\rm tot}(z_{\rm f})$. In all cases, the final result depends not only on the size of the generated asymmetry, but also on how much of it survives the subsequent washout in different flavor directions.

For Model B, BP1 and both points of Model C, the dominant contribution comes from the close-mass $N_2$-$N_3$ pair, while $N_1$ is much lighter. The corresponding CP asymmetries in Table~\ref{FourBPs}, especially in the $\tau$ flavor $\epsilon_{2,3\tau}$, generate a sizeable asymmetry $N_{B-L}^{\tau\tau}$ at early times, while the $\mu$ component $\epsilon_{2,3\mu}$ and $N_{B-L}^{\mu\mu}$ carries the opposite sign and partially compensates the $N_{B-L}^{\rm tot}$ asymmetry, as can be clearly seen from Fig.~\ref{fig:NBL_evo}. This is followed by the later $N_i$ washout stage, which reduces but does not erase all the previously generated asymmetry. The sign flip of the $e$ component $N_{B-L}^{ee}$ visible in Fig.~\ref{fig:NBL_evo} is a genuine density matrix effect. It is sourced not only by the tiny diagonal $\epsilon_{ie}$, but also by the redistribution induced by off-diagonal flavor correlations during coherent flavor evolution and decoherence. The final asymmetry $N_{B-L}^{\rm f}$ nevertheless remains positive and of the correct order.

The Model B, BP2 follows a different behavior, as shown in the second row of Fig.~\ref{fig:NBL_evo}. The relevant close-mass pair is $N_1$-$N_2$, with $M_{1,2}$ around $10^7~{\rm GeV}$, and the dominant generation of asymmetry and washout takes place in the same temperature range. In the $N_{1,2}$ decay and washout region, the generated $\mu$ asymmetry $\epsilon_{1,2\mu}$ and $\tau$ asymmetry $\epsilon_{1,2\tau}$ are nearly equal in size, but both undergo relatively strong washout. By contrast, the $e$ asymmetry $\epsilon_{1,2e}$ has the opposite sign and a smaller magnitude, while being subject to much weaker washout. Although the initially produced flavored CP asymmetries are smaller than in the other three cases, this flavor structure still allows a viable final asymmetry. The evolution in Fig.~\ref{fig:NBL_evo} therefore reflects a more direct competition between production and washout.

\subsubsection{Predictive Effective Neutrino Mass of $0\nu\beta\beta$ Decay} 
A useful probe of the low energy implications of the viable leptogenesis parameter space is neutrinoless double beta ($0\nu\beta\beta$) decay. The $0\nu\beta\beta$ decay is a hypothetical lepton number violating process in which a nucleus decays with the emission of two electrons but no neutrinos. Its observation would establish the Majorana nature of neutrinos and provide an important low energy probe of the parameter space considered here. 

In the models we considered, the contribution from heavy RH neutrinos is negligible because their mixing with the light neutrinos is strongly suppressed. The decay is therefore dominated by the standard light neutrino exchange contribution, characterized by the effective neutrino mass 
\begin{align}
\langle m_{ee}\rangle
\equiv\sum_{i=1}^{3}U_{ei}^{\,2}m_i
=c_{13}^{2}c_{12}^{2}m_{1}+s_{12}^{2}c_{13}^{2}m_{2}e^{i\alpha/2}
+s_{13}^{2}m_{3}e^{i(\beta-\delta_{\rm CP})/2}~.
\end{align}
This quantity depends on the four scanned input parameters $m_{\rm lightest}$, $\delta_{\rm CP}$, $\alpha$, and $\beta$. At present, the strongest experimental constraint comes from KamLAND-Zen 800~\cite{KamLAND-Zen:2024eml}, based on the $^{136}\mathrm{Xe}$ isotope, which sets a lower bound of $3.8\times10^{26}\mathrm{yr}$ on the half-life. Allowing for the sizable uncertainties in the nuclear matrix elements, this translates into the bound $|\langle m_{ee}\rangle|<28-122~\rm{meV}$. A good review on $0\nu\beta\beta$ decay experimental status and prospects can be found in Ref.~\cite{Giuliani:2026syi}.

\begin{figure}[t]
    \centering
    \includegraphics[width=0.95\textwidth]{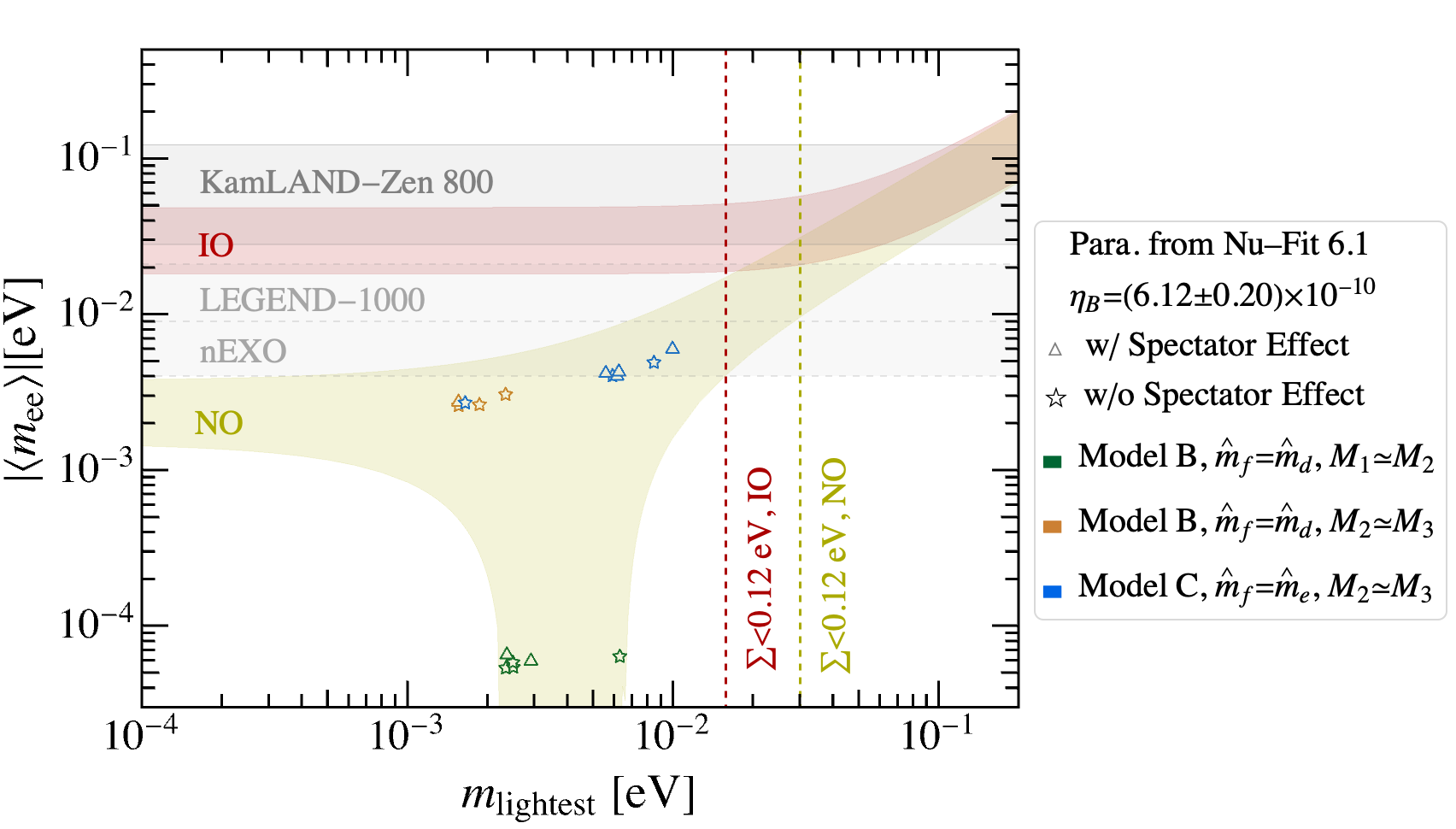}
    \caption{Effective neutrino mass $|\langle m_{ee}\rangle|$ as a function of the lightest neutrino mass $m_{\rm lightest}$. The points shown can give acceptable baryon asymmetry around $\eta_{B}=(6.12\pm 0.20)\times 10^{-10}$. For more details, see the text.}
    \label{0vbb}
\end{figure}

Figure~\ref{0vbb} presents the effective neutrino mass $|\langle m_{ee}\rangle|$ as a function of the lightest neutrino mass $m_{\rm lightest}$ for parameter points that reproduce the observed baryon asymmetry, $\eta_B=(6.12\pm0.20)\times10^{-10}$. Stars and triangles denote the cases without and with the instantaneous spectator switch, respectively. Green points correspond to Model B with the $N_1$-$N_2$ close-mass pair, orange points to Model B with the $N_2$-$N_3$ close-mass pair, and blue points to Model C with the $N_2$-$N_3$ close-mass pair. The red and yellow shaded regions represent the inverted and normal ordering bands, while the vertical dashed lines indicate the cosmological upper limits on the sum of neutrino masses $\sum m_{i}<0.12~\rm eV$~\cite{Planck:2018vyg,Allali:2024aiv} with different orderings. The gray horizontal bands mark the current and projected experimental sensitivities.

Figure~\ref{0vbb} shows that all viable points lie below the current KamLAND-Zen 800 sensitivity. Present $0\nu\beta\beta$ decay searches therefore do not probe the parameter space compatible with successful leptogenesis. Part of the Model C points fall within the target sensitivity of future experiments at the sub-$10~\mathrm{meV}$ level, including LEGEND-1000~\cite{LEGEND:2021bnm}, nEXO~\cite{nEXO:2021ujk}, JUNO 50 tons~\cite{Zhao:2016brs}, and CUPID-1T~\cite{CUPID:2022wpt}. Once the theoretical uncertainty in leptogenesis is taken into account, certain Model B points with the close-mass pair $M_2\simeq M_3$ may also enter the projected reach. By contrast, Model B points with the close-mass pair $M_1\simeq M_2$ remain in the sub-meV region, well below the sensitivity expected from future searches.

For completeness, it is useful to comment on the effective electron neutrino mass relevant for beta decay experiments $m_{\beta}=\sqrt{\sum_i |U_{ei}|^2 m_i^2}$. Current direct searches remain sensitive only to much larger values. The experiment KATRIN~\cite{KATRIN:2024cdt} reaches the level of $m_\beta\sim 0.2~\mathrm{eV}$, while Project 8~\cite{Project8:2017nal} aims at a future sensitivity of $m_\beta<0.04~\mathrm{eV}$. In contrast to $|\langle m_{ee}\rangle|$, the viable points in Models B and C lead to very similar values of $m_\beta$, typically around $0.01~\mathrm{eV}$, however, still below the projected sensitivity of future experiments. Another quantity of interest is the off-diagonal effective mass parameter $\langle m_{e\mu}\rangle \equiv \sum_i U_{ei} U_{\mu i} m_i$. Typical viable points give $|\langle m_{e\mu}\rangle|\sim 0.006-0.01~\mathrm{eV}$. Such values remain far below the sensitivity implied by current limits on $\mu^- \to e^+$ conversion, such as the SINDRUM constraint~\cite{SINDRUMII:1998mwd}, and also stay well outside the expected reach of near future experiments including Mu2e~\cite{Mu2e:2014fns} and COMET~\cite{COMET:2018auw}.

\section{Conclusions and Prospects}
\label{sec:conclusion}

This work studied thermal leptogenesis in a class of predictive Type-I seesaw models, where the neutrino Dirac mass matrix $m_{\nu}^D$ follows one of the running charged fermion mass matrices. The heavy right-handed Majorana neutrino mass matrix $M_{R}$ can then be reconstructed from low energy neutrino parameters through the seesaw relation, which greatly reduces the freedom of the high energy neutrino sector. Apart from the measured oscillation angles and mass squared differences, the remaining free parameters consist of the lightest neutrino mass $m_{\rm lightest}$, the Dirac CP phase $\delta_{CP}$ and two Majorana phases $\alpha,\beta$. As a result, leptogenesis becomes highly predictive and can be directly connected with low energy neutrino observables.

The calculation of the baryon asymmetry used the density matrix Boltzmann equations, including spectator effects. A broad random scan, followed by local refinements around regions with two close heavy neutrino masses, identified the viable parameter space. Within the parameter ranges considered, successful leptogenesis occurs only for normal ordering of the light neutrino masses. The scan found no viable region for inverted ordering. Among the three predictive models, Model A, which corresponds to the up-quark Dirac spectrum, cannot reproduce the observed baryon asymmetry. Even after a refined scan around the $|M_i-M_j|/M_{i}<10^{-4}$ region, the largest value of $\eta_B$ remains far below the observed value. This mainly results from a small CP asymmetry combined with sizable washout.
Models B and C, in contrast, can reproduce the observed value of $\eta_B$. The successful points occupy restricted regions of parameter space, typically with
$m_{\rm lightest}\in [10^{-3},10^{-2}]~ {\rm eV}$, and require two heavy neutrinos with nearby masses $|M_i-M_j|/M_{i}<10^{-3}$. In Model B, viable solutions appear in both the $M_1\simeq M_2$ and $M_2\simeq M_3$ regions. In Model C, successful points mainly occur in the $M_2\simeq M_3$ region. Although the nearby heavy neutrino masses enhance the generated asymmetry, the viable benchmark points do not enter the conventional resonant leptogenesis regime. 
The viable regions show clear correlations with the low energy CP phases. The Dirac phase $\delta_{\rm CP}$ alone does not suffice to generate the observed baryon asymmetry. Nonzero Majorana phases play an essential role. For the viable $M_2\simeq M_3$ regions of Models B and C, the phase $\alpha$ concentrates around $\pi/2$, while $\beta$ mostly lies in the interval $[\pi/4,3\pi/4]$. The phase $\delta_{\rm CP}$ tends to avoid part of the interval around $[3\pi/4,5\pi/4]$. The $M_1\simeq M_2$ region of Model B shows an even stronger concentration of $\alpha$ around $\pi/2$, together with broader allowed ranges for $\delta_{\rm CP}$ and $\beta$. These correlations provide a characteristic consequence of reconstructing $M_R$ from low energy inputs.

The analysis also examined the impact of spectator effects. Besides the standard instantaneous transition, a finite rate treatment was considered, where the muon and strange Yukawa interactions enter equilibrium gradually. For the benchmark points, the induced shift in the final baryon asymmetry ranges from $0.04\%$ to $10.66\%$. The effect remains negligible when the dominant asymmetry arises well below the equilibration temperature of the muon and strange Yukawa interactions, as in the $M_1\simeq M_2$ benchmark of Model B. It becomes more visible for the $M_2\simeq M_3$ benchmarks, where leptogenesis occurs around temperatures of order $10^9$ GeV. In all cases, the finite rate treatment changes the numerical value of the asymmetry but does not modify the qualitative viability of the parameter regions.

The viable parameter space also leads to predictions for neutrinoless double beta decay. In the models considered here, the heavy neutrino contribution remains negligible, so the standard light neutrino exchange dominates. All viable points lie below the current KamLAND-Zen 800 sensitivity. Future experiments with sub-10 meV sensitivity, such as LEGEND-1000, nEXO, JUNO 50 tons and CUPID, can probe part of the Model C parameter space. After including the theoretical uncertainty in the leptogenesis calculation, some Model B points in the $M_2\simeq M_3$ region may also become testable. By contrast, the Model B solutions in the $M_1\simeq M_2$ region predict $|\langle m_{ee}\rangle|$ in the sub-meV range and remain beyond the reach of planned searches.

Several directions deserve further study. A more complete estimate of theoretical uncertainties requires a systematic treatment of oscillation parameter inputs, running charged fermion masses, and matching conditions in different effective theories. Additional thermal corrections and scattering processes could also improve the leptogenesis calculation, together with a more complete finite rate description of spectator processes. In the future, improved measurements of the leptonic CP phase, stronger cosmological bounds on the absolute neutrino mass scale, and neutrinoless double beta decay searches will further test the viable regions identified in this work.

\section*{Acknowledgements}
Authors sorted by last name. We thank Pasquale Di Bari, Shao-Long Chen and Yong Du for their useful suggestions and comments. RRV thanks his co-authors for their kind hospitality at TDLI while part of this work was done. This work is supported in part by the National Natural Science Foundation of China (Nos. 12090064, 12375088, W2441004, 12547133), the Australian Research Council through the ARC Centre of Excellence for Dark Matter Particle Physics (CE200100008), and the China Postdoctoral Science Foundation (No. 2026T005LZD).

\bibliographystyle{JHEP}
\bibliography{references}

\end{document}